\documentclass[usegraphicx,useAMS,usenatbib]{mn2e}
\usepackage{eso-pic}

\usepackage[T1]{fontenc}
\usepackage{verbatim}
\usepackage[normalem]{ulem}
\usepackage{amsfonts,amsmath,amssymb}
\usepackage{times}
\usepackage[italic]{mathastext}
\usepackage{enumitem}
\usepackage[colorlinks,linkcolor=blue,citecolor=blue,urlcolor=blue,breaklinks]{hyperref}
\usepackage{color}
\usepackage{upgreek}
\usepackage{flushend}
\usepackage{lineno}
\usepackage{hyperref}
\usepackage{siunitx}
\usepackage{mathrsfs}
\usepackage[normalem]{ulem}

\newcommand{\redmapper}{redMaPPer}

\newcommand{\bpz}{\textsc{bpz}}

\newcommand{\metacal}{\textsc{metacalibration}}

\newcommand{\calB}{{\cal B}}

\usepackage{todonotes}

\newcommand{\addgals}{\textsc{addgals}}

\AddToShipoutPictureBG*{%
  \AtPageUpperLeft{%
    \hspace{4\paperwidth}%
    \raisebox{-4.8\baselineskip}{%
      \makebox[0pt][l]{\textnormal{DES-2018-0341}}
}}}%

\AddToShipoutPictureBG*{%
  \AtPageUpperLeft{%
    \hspace{4\paperwidth}%
    \raisebox{-5.8\baselineskip}{%
      \makebox[0pt][l]{\textnormal{FERMILAB-PUB-18-681-AE}}
}}}%

\title[Validation of weak lensing cluster member contamination estimates from P(z) decomposition in DES Y1 data]{Dark Energy Survey Year 1 results: Validation of weak lensing cluster member contamination estimates from P(z) decomposition}

\author[Varga et al.]{
\parbox{\textwidth}{
\Large
T.~N.~Varga,$^{1,2}$\thanks{corresponding author: \href{mailto:t.varga@physik.lmu.de}{t.varga@physik.lmu.de}}
J.~DeRose,$^{3,4,5}$
D.~Gruen,$^{3,5,4}$
T.~McClintock,$^{6}$
S.~Seitz,$^{1,2}$
E.~Rozo,$^{7}$
M.~Costanzi,$^{2}$
B.~Hoyle,$^{1,2}$
N.~MacCrann,$^{8,9}$
A.~A.~Plazas,$^{10}$
E.~S.~Rykoff,$^{4,5}$
M.~Simet,$^{11,12}$
A.~von der Linden,$^{13}$
R.~H.~Wechsler,$^{3,4,5}$
J.~Annis,$^{14}$
S.~Avila,$^{15}$
E.~Bertin,$^{16,17}$
D.~Brooks,$^{18}$
E.~Buckley-Geer,$^{14}$
D.~L.~Burke,$^{4,5}$
A.~Carnero~Rosell,$^{19,20}$
M.~Carrasco~Kind,$^{21,22}$
J.~Carretero,$^{23}$
C.~E.~Cunha,$^{4}$
C.~B.~D'Andrea,$^{24}$
L.~N.~da Costa,$^{20,25}$
J.~De~Vicente,$^{19}$
S.~Desai,$^{26}$
H.~T.~Diehl,$^{14}$
J.~P.~Dietrich,$^{27,28}$
P.~Doel,$^{18}$
A.~E.~Evrard,$^{29,30}$
B.~Flaugher,$^{14}$
P.~Fosalba,$^{31,32}$
J.~Frieman,$^{14,33}$
J.~Garc\'ia-Bellido,$^{34}$
E.~Gaztanaga,$^{31,32}$
D.~W.~Gerdes,$^{29,30}$
R.~A.~Gruendl,$^{21,22}$
J.~Gschwend,$^{20,25}$
G.~Gutierrez,$^{14}$
W.~G.~Hartley,$^{18,35}$
D.~L.~Hollowood,$^{36}$
K.~Honscheid,$^{8,9}$
D.~J.~James,$^{37}$
T.~Jeltema,$^{36}$
K.~Kuehn,$^{38}$
N.~Kuropatkin,$^{14}$
M.~Lima,$^{39,20}$
M.~A.~G.~Maia,$^{20,25}$
M.~March,$^{24}$
J.~L.~Marshall,$^{40}$
P.~Melchior,$^{41}$
F.~Menanteau,$^{21,22}$
C.~J.~Miller,$^{29,30}$
R.~Miquel,$^{42,23}$
R.~L.~C.~Ogando,$^{20,25}$
A.~K.~Romer,$^{43}$
E.~Sanchez,$^{19}$
V.~Scarpine,$^{14}$
M.~Schubnell,$^{30}$
S.~Serrano,$^{31,32}$
I.~Sevilla-Noarbe,$^{19}$
M.~Smith,$^{44}$
F.~Sobreira,$^{45,20}$
E.~Suchyta,$^{46}$
M.~E.~C.~Swanson,$^{22}$
G.~Tarle,$^{30}$
D.~Thomas,$^{15}$
D.~L.~Tucker,$^{14}$
and Y.~Zhang$^{14}$
\begin{center} (DES Collaboration) \end{center}
\bigskip
\small{{\em Author affiliations are listed at the end of this paper.}}
}}

\begin{document}
\date{\today}
\pagerange{\pageref{firstpage}--\pageref{lastpage}}
\pubyear{2017}
\maketitle
\label{firstpage}

\begin{abstract}
Weak lensing source galaxy catalogs used in estimating the masses of galaxy clusters can be heavily contaminated by cluster members, prohibiting accurate mass calibration. In this study we test the performance of an estimator for the extent of cluster member contamination based on decomposing the photometric redshift $P(z)$ of source galaxies into contaminating and background components.
We perform a full scale mock analysis on a simulated sky survey approximately mirroring the observational properties of the Dark Energy Survey Year One observations (DES Y1), and find excellent agreement between the true number profile of contaminating cluster member galaxies in the simulation and the estimated one. We further apply the method to estimate the  cluster member contamination for the DES Y1 redMaPPer cluster mass calibration analysis, and compare the results to an alternative approach based on the angular correlation of weak lensing source galaxies. We find indications that the correlation based estimates are biased by the selection of the weak lensing sources in the cluster vicinity, which does not strongly impact the $P(z)$ decomposition method.  Collectively, these benchmarks demonstrate the strength of the $P(z)$ decomposition method in alleviating membership contamination and enabling highly accurate cluster weak lensing studies without broad exclusion of source galaxies, thereby improving the total constraining power of cluster mass calibration via weak lensing.
\end{abstract}


\section{Introduction}
\label{sec:introduction}
Galaxy clusters trace the highest peaks of the cosmic density field and their abundance and distribution constitutes a powerful cosmological probe \citep{Allen2011, CosmicVisions16}. This mode of inference poses two major tasks: detecting galaxy clusters from observational data, and defining a  mass--observable relation (MOR) to compare the observed cluster abundances with the predicted ones. 
The efficient pathways to identify galaxy clusters differ between the available wavelengths and targeted redshifts ranges: 
In optical wavelengths and low-redshifts ($z < 1$) clusters can be detected as an overdensity of quenched, red, early type galaxies \citep{KoesterAlgorithm07, Rykoff2014_RM1}, while in other wavelengths they can be identified through the X-ray emission \citep{ROSAT, Mantz2010} or through the Sunyaev-Zeldovich effect \citep{SZ70, SZ72,Bleem2015} induced by the presence of hot intra-cluster gas. While these methods are suitable to detect clusters, they do not provide a direct measure of their masses. 
The MOR must be calibrated  using additional information.

The best method for calibrating cluster masses today is via weak gravitational lensing, as it is directly sensitive to the gravitational potential. For this reason weak lensing cluster mass calibration studies \citep{WtGI,vonderLinden14, Applegate2014, Hoekstra2015, Mantz2015, okabesmith16, Battaglia2016, rmsva, Simet2017, Murata2017, Dietrich2017, rmy1} have become a necessary component of cluster cosmology analyses.
Weak lensing mass estimates carry their own set of uncertainties, both systematic and statistical. It is expected that to fully utilize the statistical power of ongoing sky surveys, the amplitude of the MOR must be calibrated with at most a few percent total uncertainty \citep{Weinberg2012}.
With the growing depth, area, and statistical power of various sky surveys the proper characterization of systematic uncertainties is becoming the highest priority. Indeed, current analyses  are systematics dominated (e.g. Table 6 of \citealt{rmy1}), meaning that to improve on the overall cosmological constraining power we have to improve our understanding of systematic errors in the mass calibration. 

One important systematic impacting weak lensing analyses is the contamination of the  source galaxy catalog with galaxies associated with the cluster. This contamination is a result of the uncertainty in photometric redshift estimates, as few-band surveys do not provide enough information to select a pure and close to complete background sample of galaxies. 
Contaminating galaxies dilute the measurement, requiring one to boost the raw signal to recover the true signal. Hence the effect is traditionally referred to as the \emph{boost factor} \citep{Sheldon04.1, Applegate2014, Hoekstra2015, Gruen2014, Simet2017, rmsva, Medezinski17, Leauthaud17}.  Conversely, when many-band photometric information is available, the contamination can be strongly reduced, but with increased observational cost \citep{Applegate2014}.

Previous studies made use of multiple approaches in characterizing cluster member contamination: \cite{Sheldon04.1} and \cite{Simet2017} estimated the boost factor profiles from the transverse correlation of source galaxies around cluster centers, while \cite{Applegate2014} and \cite{Medezinski17,HSC18} utilized the color information in a ``color-cut'' method. \cite{Gruen2014} and \cite{Dietrich2017}  estimated the correction factor from  decomposing the source population into a cluster and background component. This latter method was expanded by \cite{rmsva} who estimated the contamination rate based on a decomposition of the  photometric redshift $P(z)$ estimates of source galaxies, which was also employed by \cite{Chang17} and \cite{Stern18}.

In this study we aim to validate the cluster member contamination estimates obtained through $P(z)$ decomposition, and provide a detailed description for the case of the DES Y1 cluster weak lensing analysis of \cite{rmy1}.
The structure of this paper is as follows: In \autoref{sec:formalism} we outline the framework and formalism of the $P(z)$ decomposition method, in \autoref{sec:method_validation} we perform tests on simulated DES-like observations  as well as actual DES data, and finally in \autoref{sec:desy1res} we present the boost factor results used in the DES Y1 redMaPPer weak lensing cluster mass calibration \citep{rmy1}.

For the DES Y1 data analysis, we assume a flat $\Lambda$CDM cosmology with $\Omega_{\rm m}=0.3$ and $H_0=70$ km s$^{-1}$ Mpc$^{-1}$, with distances defined in physical coordinates, rather than comoving. The DES-like mock observations assume a flat $\Lambda CDM$ cosmology with $\Omega_{\rm m} = 0.286$, $H_0=70$ km s$^{-1}$ Mpc$^{-1}$, $\Omega_{\rm b} = 0.047$, $n_s = 0.96$, and $\sigma_8 = 0.82$.

\section{P(z) decomposition formalism}
\label{sec:formalism}

Our aim is to estimate the cluster member contamination affecting weak lensing measurements. With an estimate of the contamination rate that has sufficiently low systematic and statistical uncertainty, we can correct for the bias in the raw weak lensing signal.

\begin{figure*}
	\includegraphics[width=\linewidth]{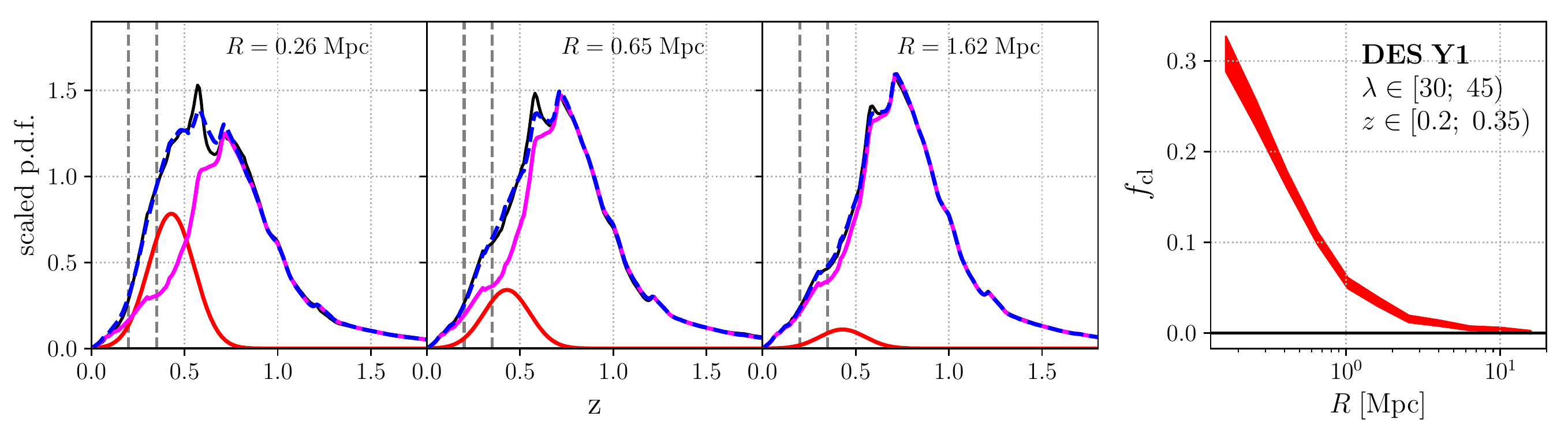}
	\caption{ \emph{(Left panels:)}  $P(z)$ decomposition at three different radial ranges for the cluster sample with richness $\lambda\in[30;\; 45)$ and redshift $z\in[0.2; 0.35)$ in DES Y1 data. \emph{Black lines:} average weighted $P(z\, |\, R)$ of source galaxies.
   \emph{Red lines:} $P(z)$ of the Gaussian contamination component scaled by the estimated cluster member contamination rate $f_\mathrm{cl}$.
    \emph{Magenta lines:} average $P(z\, |\, \mathrm{field})$ scaled by $1 - f_\mathrm{cl}$.
    \emph{Blue dashed:} model $P(z)$ calculated from  the sum of the magenta and red lines. The vertical dashed lines indicate the redshift range of galaxy clusters in the cluster selection.
	\emph{(Right panel:)} the cluster member contamination rate $f_\mathrm{cl}$ profile calculated from the decomposition presented on the left panels: the red shaded range corresponds to the amplitudes of the Gaussian components at each radial range. The width of the shaded area indicates the $1\sigma$ uncertainty region. }
    \label{fig:boost_profiles_data}
\end{figure*}

The present approach infers the fraction of contaminating cluster member galaxies $f_\mathrm{cl}$ from the photometric redshift $P(z)$ probability distribution function (p.d.f.) of the appropriately selected and weighted source galaxies. By comparing the $P(z)$ of sources near clusters with the $P(z)$ of galaxies in field lines-of-sight we identify a feature indicative of the presence of cluster galaxies, illustrated by the red curve in the left panels of \autoref{fig:boost_profiles_data}. The relative strength of this feature at different radii is taken as a tracer of the cluster member contamination rate profile $f_\mathrm{cl}(R)$, shown on the right panel of \autoref{fig:boost_profiles_data}.



\subsection{Weak lensing formalism}
\label{sec:wlformalism}
Weak lensing analyses of galaxy clusters rely on a large sample of background \emph{source} galaxies. The images of these background source galaxies are distorted due to the gravitational potential of the lens (the galaxy cluster), and thus can be used to trace the underlying matter distribution of the lens \citep{Bartelmann01.1}

In most scenarios a catalog of background source  galaxies is constructed from optical imaging data, and thus their exact distances are not known. To remedy this, photometric redshift algorithms are employed to provide an estimate of their redshifts. Such methods involve large uncertainty for individual galaxies, and potential bias for the ensemble, due to the limited information available \citep{Y1pz, KIDS450}. The uncertainty of photometric redshifts mean that the background source catalog can only be defined in approximate terms, such that it may also include foreground galaxies, and galaxies which are at the lens redshift. 
In contrast, the redshifts of galaxy clusters are typically known with very high precision either from spectroscopic follow up, or from the ensemble photometric redshift estimates of their red cluster member population \citep{Rykoff2016, rmy1}. In this analysis we neglect any uncertainty in the cluster redshift $z_\mathrm{lens}$, and consider that the uncertainty in the redshift of source galaxies $z_\mathrm{src}$ is captured in their $P(z_\mathrm{src})$ p.d.f.

Background galaxies ($z_\mathrm{src} > z_\mathrm{lens}$) at different redshifts (distances) contribute to the lensing signal with different amplitudes. This is characterized by the inverse of the \emph{critical surface density}:
\begin{equation}
\Sigma_\mathrm{crit}(z_\mathrm{lens}, \, z_\mathrm{src}) = \frac{c^2}{4 \pi G} \frac{D_\mathrm{s}(z_\mathrm{src})}{D_\mathrm{l}(z_\mathrm{lens}) \, D_\mathrm{ls}(z_\mathrm{lens}, z_\mathrm{src})}\,,
\label{eq:sigmacrit}
\end{equation}
where $D_\mathrm{s}$, $D_\mathrm{l}$ and $D_\mathrm{ls}$ denote \emph{angular diameter} distances to the source galaxy, the lens, and between the lens and the source respectively.

In a cluster weak lensing scenario the quantity of interest is the average tangential component of the reduced gravitational shear $g_\mathrm{T} = \gamma_\mathrm{T} / (1 - \kappa)$ where $\gamma$ is the weak lensing shear and $\kappa$ is the convergence. This is estimated from the shapes and alignments of the source galaxies through the ellipticity measure $\mathbf{e}$, where we assume that  $\langle \mathbf{e}  \rangle \approx \langle \mathbf{g} \rangle$. The shear signal is related to the \emph{excess surface mass density} $\Delta\Sigma$, expressible from the physical mass distribution of the lens system via:
\begin{equation}
\gamma_\mathrm{T}(R) = \frac{\overline{\Sigma}(<R) - \overline{\Sigma}(R) }{\Sigma_\mathrm{crit}} = \frac{\Delta\Sigma(R)}{\Sigma_\mathrm{crit}}\,,
\label{eq:deltasigma}
\end{equation}
and the convergence is defined as 
\begin{equation}
\kappa(R) = \overline{\Sigma}(R)\, /\, \Sigma_\mathrm{crit}\,,
\label{eq:kappa}
\end{equation} 
where $R$ is the projected separation from the lens, $\overline{\Sigma}(<R)$ and $\overline{\Sigma}(R)$ refer to the average surface mass density within radius $R$, and at radius $R$, respectively.

Following \cite{Sheldon04.1} we define the maximum likelihood estimator for the stacked excess surface mass density of multiple clusters:
\begin{equation}
\label{eq:ideal_estimator}
\widetilde{\Delta\Sigma} = \frac{ \sum\limits_{i}^\mathrm{src} \sum\limits_{j}^\mathrm{lens} \; w_{i,j} \; \left. e^\mathrm{T}_{i,j} \middle/ \langle \Sigma_\mathrm{crit}^{-1} \rangle_{i,j} \right.}{ \sum\limits_{i}^\mathrm{src} \sum\limits_{j}^\mathrm{lens} \; w_{i,j}} =   \frac{\sum\limits_{i}^\mathrm{src} \sum\limits_{j}^\mathrm{lens}  \; w_{i,j} \; \Delta\Sigma_{i,j}}{ \sum\limits_{i}^\mathrm{src}  \sum\limits_{j}^\mathrm{lens} \; w_{i,j}}\,,
\end{equation}
with weights
\begin{equation}
w_{i,j} = \left. \langle  \Sigma_\mathrm{crit}^{-1}\rangle_{i,j}^{2} \middle/  \sigma_{e, \,i}^2 \right. \,,
\end{equation}
Here $e^\mathrm{T}_{i,j}$ corresponds to the tangential component of the  \emph{estimated ellipticity} $e$ of the $i$-th source galaxy relative to the $j$-th lens, $\sigma_{e,i}^2$ is the variance of the estimated shape for galaxy $i$, and
\begin{equation}
\langle \Sigma_\mathrm{crit}^{-1}\rangle_{i,j} = \int \mathrm{d} z_\mathrm{src} \; P_i(z_\mathrm{src}) \;\Sigma_{\mathrm{crit}, \, i,j}^{-1}(z_\mathrm{lens}, z_\mathrm{src})\,
\end{equation}
is defined as the \emph{effective} inverse critical surface density for source-lens pair $i,j$.\footnote{We estimate $\langle \Sigma_\mathrm{crit}^{-1}\rangle_{i,j}$, as $\langle \Sigma_\mathrm{crit}\rangle_{i,j}$ is less numerically stable.}. We note that $\Delta\Sigma$ relates to $\gamma$, however the distortion of source galaxies is determined by $g$, thus the effect of magnification must be accounted for during the modeling of the measured lensing signal.

\label{sec:pzestimates}
The $\widetilde{\Delta\Sigma}$ estimator defined in \autoref{eq:ideal_estimator} is unbiased if the redshift p.d.f. $P(z_\mathrm{src})$ is estimated correctly.  
In general, intrinsic bias in the photometric redshift estimates would also bias the weak lensing measurement\footnote{Depending on the form of the lensing estimator, it is possible for the redshift estimate to be biased in a way such that the lensing estimator is still unbiased. This however does not hold for arbitrary photo-z bias.}, while uncertainty alone can be propagated self-consistently \citep{Applegate2014}. For this reason, weak lensing surveys spend great effort calibrating photometric redshifts for their weak lensing source galaxy catalogs \citep{Kelly2014, Y1pz,KIDS450, HSCPz}. In case the photometric redshifts are biased, it requires additional work and loss of constraining power to ensure that the photo-z bias is appropriately propagated into the systematic error budget of the final scientific results \citep{Y13x2}.




The photometric redshift $P(z)$ assigned to a galaxy is determined by its observed properties and also by our prior knowledge about its likely redshifts. To reach good photo-z performance (e.g. low bias) the prior should be strongly dependent on the selection function. Defining this in practice requires a reference sample of galaxies for which the mapping between redshift and observed properties is known, and is representative of the target galaxy selection. Hence when redshift estimates calibrated with one selection are used together with a significantly different selection during the science analysis, they are no longer guaranteed to retain their fiducial performance. \citep{BonnettPhotoz2015, Y1pz}. 


\subsection{Boost factor formalism}
\label{sec:boost_formalism}
\label{sec:origins}

Galaxy clusters represent a large overdensity of physically associated galaxies, consisting both of actual cluster member galaxies, and also the galaxies inhabiting correlated structures. Thus cluster lines-of-sight are different from the average line-of-sight: The galaxy overdensity is concentrated at a tight peak in redshift, much narrower than the resolution of photometric redshift estimates of individual galaxies. In the transverse direction, the number density of the cluster-related galaxy population decreases as one moves away from the cluster center, where different galaxy populations (e.g. red and blue cluster galaxies) follow different radial profiles \citep{Navarro96.1, Rykoff2014_RM1}.

Observationally, the weak lensing source galaxy catalog is built from galaxies selected according to morphological, photometric and spatial selection criteria \citep{Y1shape,KIDS450, HSCshape}, and galaxies associated with the targeted clusters also enter the catalog if they satisfy those criteria. Ideally cluster galaxies would be excluded \citep[e.g. as in][]{Schrabbak2016} since they are at the lens redshift and carry no lensing signal. However in a wide field survey, priors used to estimate photometric redshifts do not account for the presence of the targeted clusters.
Consequently the redshift estimates can be greatly biased, and allow for cluster galaxies leaking into the source catalog with non-zero weights. The redshift bias may further depend on galaxy type, resulting in different rates of contamination by different populations of cluster galaxies. Hence defining a high purity background sample of sufficient volume may not be possible. We note that depending on the radial separations, the contamination can originate from both the targeted galaxy clusters and also from galaxies in the correlated matter structures. For reasons of brevity we refer to both of these sources as cluster member contamination, as they can not be disentangled based purely on available redshift information.

To quantify the required boost factor correction we need to consider the impact of contamination on the $\widetilde{\Delta\Sigma}$ estimator defined in \autoref{eq:ideal_estimator}. Following the method developed in \cite{Gruen2014} and extended in \cite{rmsva}, we assume a model for the \emph{true} line-of-sight distribution of source galaxies selected during the measurement as a combination of two terms: a cluster galaxy component which is effectively a Dirac-$\delta$ function located at $z_\mathrm{clust}$, and a non-cluster or background component taken to be the lensing weighted distribution of source galaxies in field lines-of-sight.

Via the above line-of-sight model, we can expand \autoref{eq:ideal_estimator} into the sum of contributions from the cluster ($\mathrm{cl}$) and background ($\mathrm{bg}$) terms:
\begin{equation}
\label{eq:split}
\begin{split}
\widetilde{\Delta\Sigma}_\mathrm{est} &= \frac{\sum\limits_{j,i\in \mathrm{cl}} w_{i,j} \Delta\Sigma_{i,j} + \sum\limits_{j,i\in \mathrm{bg}} w_{i,j} \Delta\Sigma_{i,j}}{\sum\limits w_{i,j}}\\
 & =\left(\frac{\sum\limits_{j,i\in \mathrm{cl}} w_{i,j}}{\sum\limits w_{i,j}}\right) \underbrace{\langle\Delta\Sigma_{i,j}\rangle_\mathrm{cl}}_{0} + \left(\frac{\sum\limits_{j,i\in \mathrm{bg}} w_{i,j}}{\sum\limits w_{i,j}}\right) \underbrace{\langle\Delta\Sigma_{i,j}\rangle_\mathrm{bg}}_{\langle\Delta\Sigma_{i,j}\rangle_\mathrm{true}}\,
 \end{split}
\end{equation}
of which $\langle\Delta\Sigma_{i,j}\rangle_\mathrm{cl}$ carries no signal, while $\langle\Delta\Sigma_{i,j}\rangle_\mathrm{bg}$ is the ``true'' signal we would estimate if there was no contamination. $\sum_{j,i\in \mathrm{cl}}$ and $\sum_{j,i\in \mathrm{bg}}$ denotes a sum over source-lens pairs with cluster members and background galaxies respectively. We furthermore define the effective contamination rate of cluster galaxies $f_\mathrm{cl}$ via 
\begin{equation}
f_{\rm cl} = \frac{\sum_{j,i\in \mathrm{cl}} w_{i,j}}{\sum_{j,i} w_{i,j}}\, \;,
\label{eq:f_clust}
\end{equation}
which we can use to express the boost correction needed to recover the true signal
\begin{equation}
\widetilde{\Delta\Sigma}_{\rm corr}(R) = \frac{\widetilde{\Delta\Sigma}(R)}{1 - f_{\rm cl}(R)}\,.
\label{eq:boost_correction}
\end{equation}
Here $\widetilde{\Delta\Sigma}$ denotes the raw measured lensing signal obtained from \autoref{eq:ideal_estimator}, $\widetilde{\Delta\Sigma}_{\rm corr}(R)$ denotes the lensing signal corrected for contamination, and $\mathcal{B}\equiv  (1 - f_{\rm cl})^{-1}$ is referred to as the \emph{boost factor}. Hence in the above framework the cluster member contamination correction for a given measurement scenario is completely characterized by the $f_\mathrm{cl}(R)$ profile.

\subsection{Estimating the contamination using P(z) decomposition}
\label{sec:boost_estimator}


We estimate the contamination rate from the available color--magnitude information of source galaxies, where, due to the overdensity of the cluster we expect that the contaminating cluster galaxies will appear as a sub-population. We follow \cite{rmsva}, and make use of the {\it observed} lens-weighted average redshift probability distribution of the sources
\begin{equation}
	P(z\,\vert\, R) = \frac{\sum_{i,j} w_{i,j}\, P_i(z, R)}{\sum_{i,j} w_{i,j}}\,,
    \label{eqn:boostpzweight}
\end{equation}
which compresses information from color--magnitude space into a probability distribution. We measure this at different projected radii $R$ around the cluster. The weights $w_{i,j}$ for each source are identical to the ones introduced in \autoref{eq:ideal_estimator}.  In this framework the estimated redshifts represent information compression from the color--magnitude space into a single $P(z)$ estimate per sample. Contaminating cluster members contribute to the average photometric redshift $P(z)$-s differently in different radial ranges. Thus by tracking the changes in the  $P(z)$ as a function of radius, we can recover an estimate of the underlying cluster member contamination.

We model the observed redshift distribution $P(z\,|\, R)$  as a combination of two terms, reflecting the cluster and background populations defined in \autoref{sec:boost_formalism}:
\begin{equation}
	\label{eq:pz_decomposition} 
	P(z\, \vert \, R) = f_\mathrm{cl}(R) \cdot P_\mathrm{memb}(z) + (1 - f_\mathrm{cl}(R)) \cdot P_\mathrm{bg}(z)\,,
\end{equation}
where $P_\mathrm{memb}(z)$ is the redshift distribution of \emph{contaminating} cluster member galaxies  and $P_\mathrm{bg}(z)$ is the distribution of background galaxies \citep{Gruen2014, rmsva}. We approximate the second term by the appropriately weighted redshift distribution of the average survey \emph{field} line-of-sight: $P_\mathrm{bg}(z) \approx P_\mathrm{field}(z)$. As an ansatz we consider $P_\mathrm{memb}(z)$ to be a Gaussian distribution. The validity of this assumption is tested in \autoref{sec:member_pz}.  
The free parameters of the decomposition are the mean and width of the Gaussian $P_\mathrm{memb}(z)$, and the contamination rate $f_{\rm cl}(R)$. 

An example for this $P(z)$ decomposition method is shown on \autoref{fig:boost_profiles_data} for the case of DES Y1 data. There, a qualitatively similar behavior is visible for the different radial bins, and the contamination increases with decreasing radius. 

\section{Method validation}
\label{sec:method_validation}
We perform a validation benchmark to test the robustness and performance of the $P(z)$ decomposition boost estimates. First, we outline the primary assumptions of the decomposition method in \autoref{sec:assumptions}, validate the method in a mock analysis scenario in \autoref{sec:buzzard_tests}, and  perform consistency tests on DES Y1 data in \autoref{sec:desy1_tests}.

\subsection{Model assumptions}
\label{sec:assumptions}
The $P(z)$ decomposition method relies on several assumptions about both the contaminating and the background galaxies which impact the efficacy of the method. Some of these we explore below, i.e. the potential intrinsic alignment of cluster galaxies (\autoref{sec:ia}), the impact of weak lensing magnification (\autoref{sec:magnification}), and the influence of blending and intra-cluster light on photometry used in the decomposition estimates (\autoref{sec:icl}). Other assumptions are tested in later sections, i.e. the Gaussian ansatz for $P_\mathrm{memb}(z)$ (\autoref{sec:member_pz}), and the influence of the chosen background model (\autoref{sec:iterative_test}). 



\subsubsection{Intrinsic alignments}
\label{sec:ia}
Contaminating galaxies physically connected to the lens system possess an \emph{intrinsic alignment} due to the tidal forces acting between them. When cluster members are included in a lensing measurement, intrinsic alignments could appear as negative tangential shear around clusters due to the preferential radial orientation of galaxies. This effect is difficult to decouple from the physical lensing signal.
Recent spectroscopic follow-up studies of \cite{Hao2011, Sifon2015} found no significant signal for preferential alignments of cluster member galaxies with respect to the cluster centers. \cite{Huang17} found significant detection only when considering a high luminosity subset of galaxies, but no detection when considering their complete galaxy sample.
For this reason  we assume that the dominant effect of contaminating cluster members is the dilution of the lensing signal.

\subsubsection{Magnification}
\label{sec:magnification}

Weak lensing magnification caused by the potential of the cluster changes the observed number density and luminosity function of background galaxies. This translates into biased photometric redshift estimates for background galaxies, which would result in the $P(z)$ of background galaxies near clusters to be different from the $P(z)$ of the similarly weighted galaxies in field lines-of-sights. \citet[][their Appendix C]{Gruen16}, however, finds, that under realistic survey assumptions biases in photometric redshift estimates due to the increased flux of magnified sources, and due to the different surface density of magnified galaxies are sub per-cent effects that partially cancel one another.


\subsubsection{Impact of blending and intra-cluster light}
\label{sec:icl}
The potential bias due to blending and source obscuration in the estimated $P(z)$-s is difficult to estimate, as it would require detailed understanding of detection and shape measurement selection probabilities, as well as the photometric transfer function of representative source galaxy samples in cluster fields \citep[see e.g.][]{Chang2015,Suchyta2016}.  To a first approximation, we expect blending to uniformly impact all source galaxies, leading to an amplitude shift in the $P(z)$ of the selected source galaxies. However given the excellent match of the cluster background population $P(z)$ and the field background population $P(z)$ at large $z\gg z_l$ (visible in \autoref{fig:boost_profiles_data}), we assume that the impact of blending is strongly subdominant. In  \autoref{sec:iterative_test} we nevertheless perform a simple consistency test for differences in the background $P(z)$.

The presence of \emph{intra-cluster light} biases the photometric redshift estimates, influencing the recovered $P(z)$-s in a manner similar to blending. However \cite[][their Appendix A]{Gruen2018_ICL} estimated the impact of this effect to be negligible for the radial scales considered in this study.

\subsection{Benchmark on the Buzzard mock observations}
\label{sec:buzzard_tests}

\begin{figure} 
    \includegraphics[width=\linewidth]{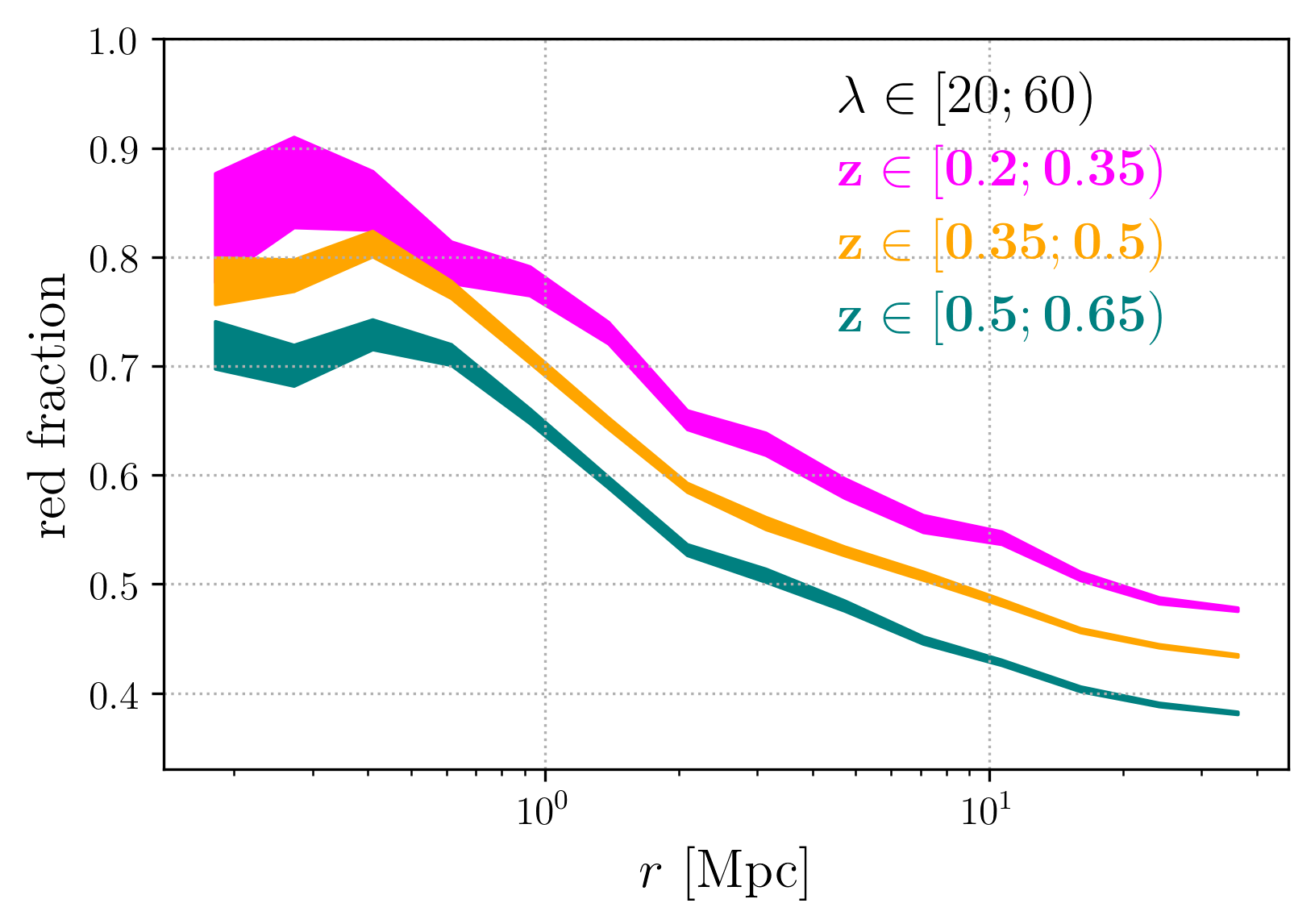} 
    \caption{Fraction of red galaxies as a function of radius around redMaPPer clusters within the \emph{Buzzard} mock observation across different redshift ranges but fixed richness. Shaded areas indicate $1\sigma$ statistical uncertainties estimated from Jackknife resampling.
	}
    \label{fig:redfraction}
\end{figure}

\begin{figure*} 
    \includegraphics[width=\linewidth]{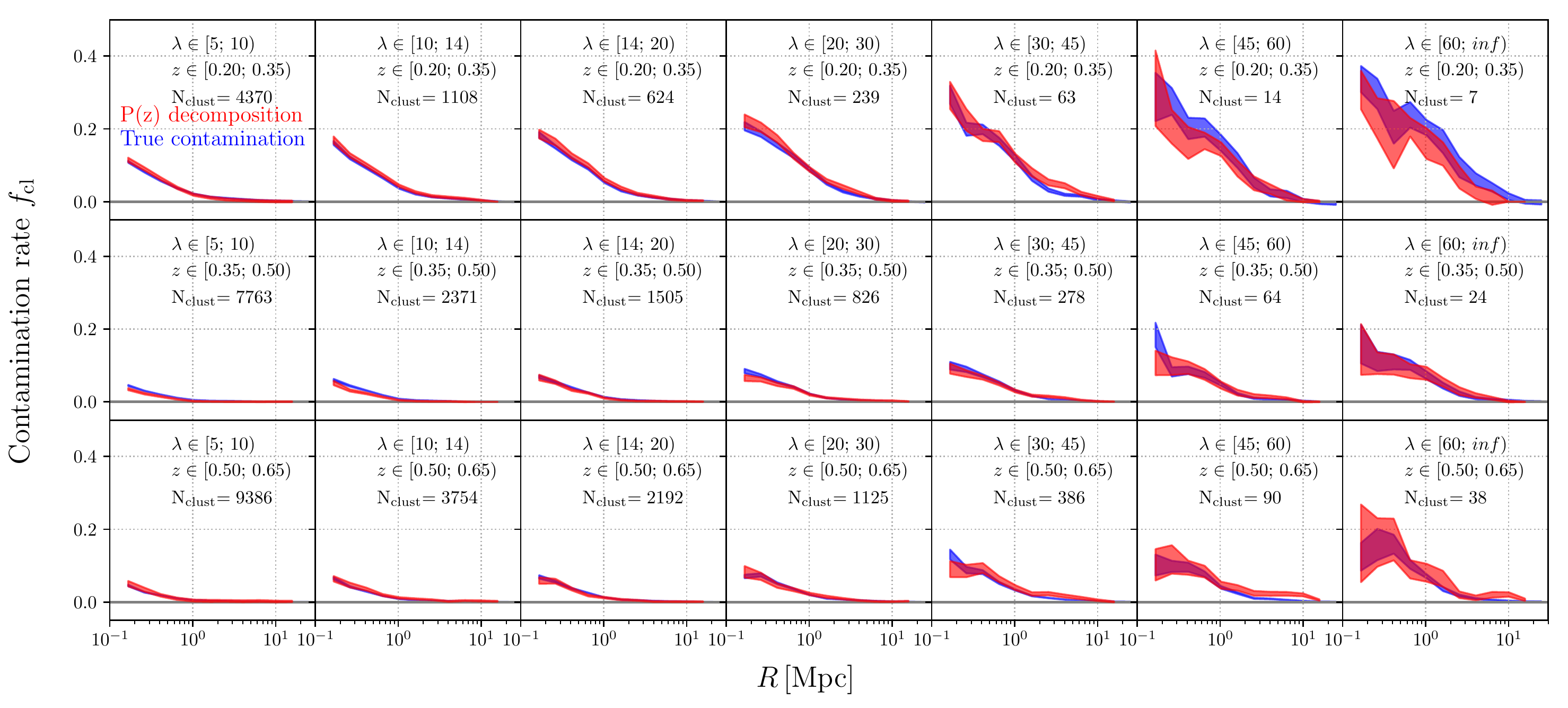}    
	\caption{
Cluster member contamination measured in the \emph{Buzzard} mock observations for the various bins of clusters in richness $\lambda$ and redshift $z$ defined in \autoref{sec:buzzard_pz}. 
\emph{Red:} contamination profiles calculated via  $P(z)$ decomposition using \bpz\ photometric redshift estimates with realistic photometric noise added to the mock galaxies.
\emph{Blue:} True contamination profiles calculated from the truth catalogs of the simulations via \autoref{eq:true_boost}.
    \label{fig:boost_profiles_buzzard}
}
\end{figure*}

We test the $P(z)$ decomposition method against the true cluster member contamination in a simulated environment, mirroring the measurement setup of \cite{rmy1}. 
In \autoref{sec:buzzard_descr} and \autoref{sec:simcluster} we introduce the simulated observations and the mock galaxy clusters. In \autoref{sec:buzzard_pz} we perform the $P(z)$-decomposition on the simulated \emph{Buzzard} data. In \autoref{sec:buzzard_true} we determine the true contamination of our photo-$z$ selected source sample. In \autoref{sec:member_pz} we test the validity of the Gaussian ansatz for $P_\mathrm{memb}(z)$. Finally, in \autoref{sec:buzzard_summary} we discuss the agreement between the true and estimated contamination rates.

\subsubsection{Buzzard simulated lightcones}
\label{sec:buzzard_descr}

The \emph{Buzzard}-suite of cosmological simulations \citep{DeRose2018} consists of mock DES Y1 catalogs generated by combining three \emph{N}-body lightcones created using L-Gadget2, a version of Gadget2 \citep{Springel05} optimized for memory efficiency. The initial conditions were set up via 2nd order Lagrangian perturbation theory using 2LPTIC \citep{Crocce06}. The lightcones were produced on the fly using simulation boxes with volumes $1050^3$, $2600^3$, and $4000^3\; (h^{-1} \mathrm{Mpc})^3$; the corresponding particle masses are $3.3 \times 10^{10}, \; 1.6 \times 10^{11}$ and $\; 5.9 \times 10^{11}\; h^{-1} \mathrm{M}_\odot$. The resulting lightcones were joined at redshifts 0.34 and 0.9, arranged such that the highest resolution simulations are used at lower redshifts.  These simulation boxes assume a flat $\Lambda CDM$ cosmology with $\Omega_{\rm m} = 0.286$, $H_0=70$ km s$^{-1}$ Mpc$^{-1}$, $\Omega_{\rm b} = 0.047$, $n_s = 0.96$, and $\sigma_8 = 0.82$. The galaxy catalogs were created by assigning galaxies to dark matter particles via the \addgals\ algorithm \citep{Wechsler2018}. \addgals\ calibrates the relation between the large scale density and the r-band absolute magnitudes of galaxies as measured using subhalo abundance matching \citep{Conroy06,Reddick13, Lehman17} in a high resolution \emph{N}-body simulation. For each simulated galaxy, SEDs are assigned from the SDSS DR7 VAGC \citep{Cooper11} by finding the galaxy in the data with the closest match in $M_{r}-\Sigma_{5}-z$ space, where $M_r$ is the galaxy's rest frame r-band absolute magnitude and $\Sigma_5$ is the distance to the 5th-nearest galaxy in projection.
Photometric noise is added in accordance with the DES Y1 depth map of \cite{Y1gold}, and \emph{g,r,i,z} fluxes in the DES filters are generated from the previously assigned SEDs.

In this study we use version 1.3 of the \emph{Buzzard} mock catalogs. Only the main ``SPT''-area of DES Y1 is simulated, and the footprint is restricted to $RA > 0$ to exclude areas where the DES coverage is more inhomogeneous. The resulting galaxy fluxes include the effect of weak lensing magnification based on ray tracing along their lines-of-sight. For the purposes of the current measurement we selected sources in a way that is meant to approximate the source selection in the DES Y1 analysis \citep{Y1shape} by applying S/N cuts following \cite{MacCrann18}. This sample is defined purely to mirror the properties  of actual DES source galaxies, and does not contain shear or photometry systematics. We then run the \bpz\ template based photometric redshift algorithm \citep{Benitez2000, coe06} on this mock catalog to obtain a $P(z)$ estimate for each source galaxy with equivalent settings as used by \cite{Y1pz} for the DES Y1 data.
Given plausible galaxy colors and identical measurement setup, we expect \bpz\ to possess similar performance in Buzzard as in the DES Y1 data

\subsubsection{Simulated galaxy clusters}
\label{sec:simcluster}

In order to obtain a simulated cluster sample similar to the one presented by \cite{rmy1}, we run the \redmapper\ algorithm \citep{Rykoff2014_RM1} on the mock galaxy catalogs with the same configuration as the real DES Y1 data. 

RedMaPPer is a red-sequence based optical matched filter cluster finder which produces an estimate on the position, the optical richness $\lambda$, and redshift of the detected clusters. This yields a cluster catalog with comparable distribution in angular position and redshift to the catalog in the DES Y1 dataset.  A catalog of reference random points is also generated, which are defined as positions and redshifts where a cluster of given richness can be detected. 
 
While the redMaPPer algorithm is sensitive only to the overdensity of red-sequence galaxies, we also test the blue galaxy content of clusters as they are expected to significantly contribute to the contamination. We calculate the fraction of red galaxies as a function of radii using a rest frame magnitude limit of $m_r >-19$. For this we take a galaxy as red if it belongs to the red sequence defined in the rest frame color -- magnitude space, which in practice corresponds to a cut of $(g-r) > 0.2 \cdot r - 0.028$.
\autoref{fig:redfraction} shows this red fraction across different redshift ranges, where we find good qualitative agreement with previous observational studies \citep[e.g][their Figure 12]{Butcher78, Hansen08}. The red fractions are different across different redshift bins with a larger blue cluster member population at higher redshifts, which is expected from the time dependence of the galaxy quenching process. 

A difference between the real and mock cluster catalogs is that clusters in the simulation appear to have a stronger redshift evolution in richness at a given halo mass relative to expectations from existing scaling relations. This fact along with the reduced simulated footprint results in a lower number of clusters in richness bins at low redshift compared to \cite{rmy1}. In addition, the DES Y1 data is deeper than the reference dataset used by the \textsc{addgals} algorithm to populate SEDs, and for this reason the mock galaxy populations and their relative abundances at faint magnitudes or high redshifts might differ from reality.
Because of this, as well as because of differences in source galaxy selection and between our real and synthetic datasets, we do not expect the cluster member contamination rates in the mock observations to be equal to our DES Y1 measurements. Nevertheless, the mocks include many qualitative aspects of the real observations, and for this reason we make use of them as a controlled environment to benchmark and validate the performance of the $P(z)$ decomposition under somewhat simplified circumstances.


\subsubsection{Decomposition results in simulated catalogs}
\label{sec:buzzard_pz}

We estimate boost factors for the \redmapper\ clusters in the mocks using an identical measurement setup as \cite{rmy1}. Hence the $\Delta\Sigma$ estimator takes the form of:
\begin{equation}
\label{eq:ds_buzzard}
\widetilde{\Delta\Sigma}=\frac{\sum \omega_{i, j} e_{\rm T;i}}{\sum \omega_{i,j}  \Sigma'^{-1}_{\rm crit;i}} \;
\end{equation}
with
\begin{equation}
	\label{eq:updated_weights}
	\omega_{i,j} \equiv \Sigma_{{\rm crit}}^{-1}\left(z_{{\rm l}_j}, \langle z_{{\rm s}_i}\rangle\right)\; \mathrm{if}\; \langle z_{{\rm s}_i}\rangle > z_{{\rm l}_j} + 0.1\,.
\end{equation}
Where $\Sigma'^{-1}_{\rm crit;i}$ is calculated at a source redshift randomly drawn from the corresponding  $P(z)$, while $\Sigma_{{\rm crit}}^{-1}\left(z_{{\rm l}_j}, \langle z_{{\rm s}_i}\rangle\right)$ represents the value at the mean redshift of the source. 
The mock galaxy catalog does not include shear biases thus we set the shear and selection responses to unity \citep[see Equation 12 of][]{rmy1}. 

Following \autoref{eq:split} and \autoref{eq:f_clust}, the contamination fraction is given by:
\begin{equation}
\label{eq:buzzard_fcl}
f_\mathrm{cl} = \frac{\sum_\mathrm{cl} \omega_{i,j}  \Sigma'^{-1}_{\rm crit;i,j}}{\sum \omega_{i,j}  \Sigma'^{-1}_{\rm crit;i,j}}\,.
\end{equation}
We select the clusters into bins of redshift $z\in[0.2;0.4)$, $[0.4;0.5)$, and $[0.5;0.65)$, and richness: $\lambda\in[5;10)$, $[10; 14)$, $[14; 20)$, $[20; 30)$, $[30; 45)$, $[45; 60)$, and $[60; \infty)$. $\widetilde{\Delta\Sigma}$ is calculated in 11 logarithmically spaced radial bins ranging from $0.2\; {\rm Mpc}$ to $30\; {\rm Mpc}$. For each cluster sample and radial range we save a representative, random subsample of source-lens pairs, and calculate the mean $P(z)$ of that source population weighted by  $\omega  \Sigma'^{-1}_{\rm crit}$.  The  estimate  on $f_\mathrm{cl}(R)$ is then found by the $P(z)$ decomposition method outlined in  \autoref{sec:boost_estimator}. For the field component we take the $P(z)$ of sources in the outermost radial bin, which we find to be identical to the weighted $P(z)$ of sources selected around random points in a series of Kolmogorov-Smirnov tests.  

The decomposition is calculated by considering all radial scales simultaneously where
we require the redshift positions and widths of the cluster components to be identical at different radial ranges. The mixing amplitudes $f_\mathrm{cl}(R)$ between the cluster and reference $P(z)$ are left free across radial bins.
Hence the inner radial scales where the contamination is stronger provide constraints about the cluster component for the outer radial ranges. The $f_\mathrm{cl}$ profile model for a cluster sample has $N_\mathrm{rbin} + 2$ free parameters, and the decomposition is performed via a least squares Levenberg-Marquardt algorithm, where the optimized quantity is the mean squared deviation between the measured $P(z\,|\,R)$ and the model prediction defined in \autoref{eq:pz_decomposition}. This boost factor calculation  algorithm is implemented in the \textsc{xpipe} python package\footnote{\url{https://github.com/vargatn/xpipe}}, which was also used by \cite{Chang17} and \cite{rmy1}, and contains an identical setup to \cite{rmsva}.

To estimate the uncertainty on the recovered $f_\mathrm{cl}(R)$ we use jackknife (JK) resampling following \cite{Efron82.1}:
\begin{equation}
\label{eq:jackknife_cov}
\mathsf{C}^{JK}_{\widetilde{f_\mathrm{cl}}} = \frac{K - 1}{K} \sum\limits_{k}^{K}\left(\widetilde{f_\mathrm{cl}}_{(k)} - \widetilde{f_\mathrm{cl}}_{(\cdot)}\right)^T \cdot \left(\widetilde{f_\mathrm{cl}}_{(k)} - \widetilde{f_\mathrm{cl}}_{(\cdot)}\right)\,,
\end{equation}
where $\widetilde{f_\mathrm{cl}}_{(\cdot)} = \frac{1}{K}\sum_k \widetilde{f_\mathrm{cl}}_{(k)}$ and $\widetilde{f_\mathrm{cl}}_{(k)}$ denotes the contamination rate estimated via \autoref{eq:buzzard_fcl}. We make use of $K=\min\lbrace 100\,;\,N_\mathrm{clust}\rbrace$ simply-connected spatial regions $\mathcal{R}_k$ for each cluster sample, defined via a spherical \textit{k-means} algorithm\footnote{\url{https://github.com/esheldon/kmeans_radec}}, and $\widetilde{f_\mathrm{cl}}_{(k)}$ is calculated from all clusters except those in region $\mathcal{R}_k$. With this method we estimate the  covariance between all radial ranges within each richness and redshift bin. 
 
The recovered contamination profiles are shown in \autoref{fig:boost_profiles_buzzard}, and are qualitatively similar to those observed in the real data.
The overall behavior is consistent with theoretical expectations: For all cluster bins the contamination rate decreases with increasing redshift, and a clear trend is apparent where richer clusters at a given redshift range produce greater contamination rates.

\subsubsection{True contamination in simulated catalogs}
\label{sec:buzzard_true}

\begin{figure}
\includegraphics[width=\linewidth]{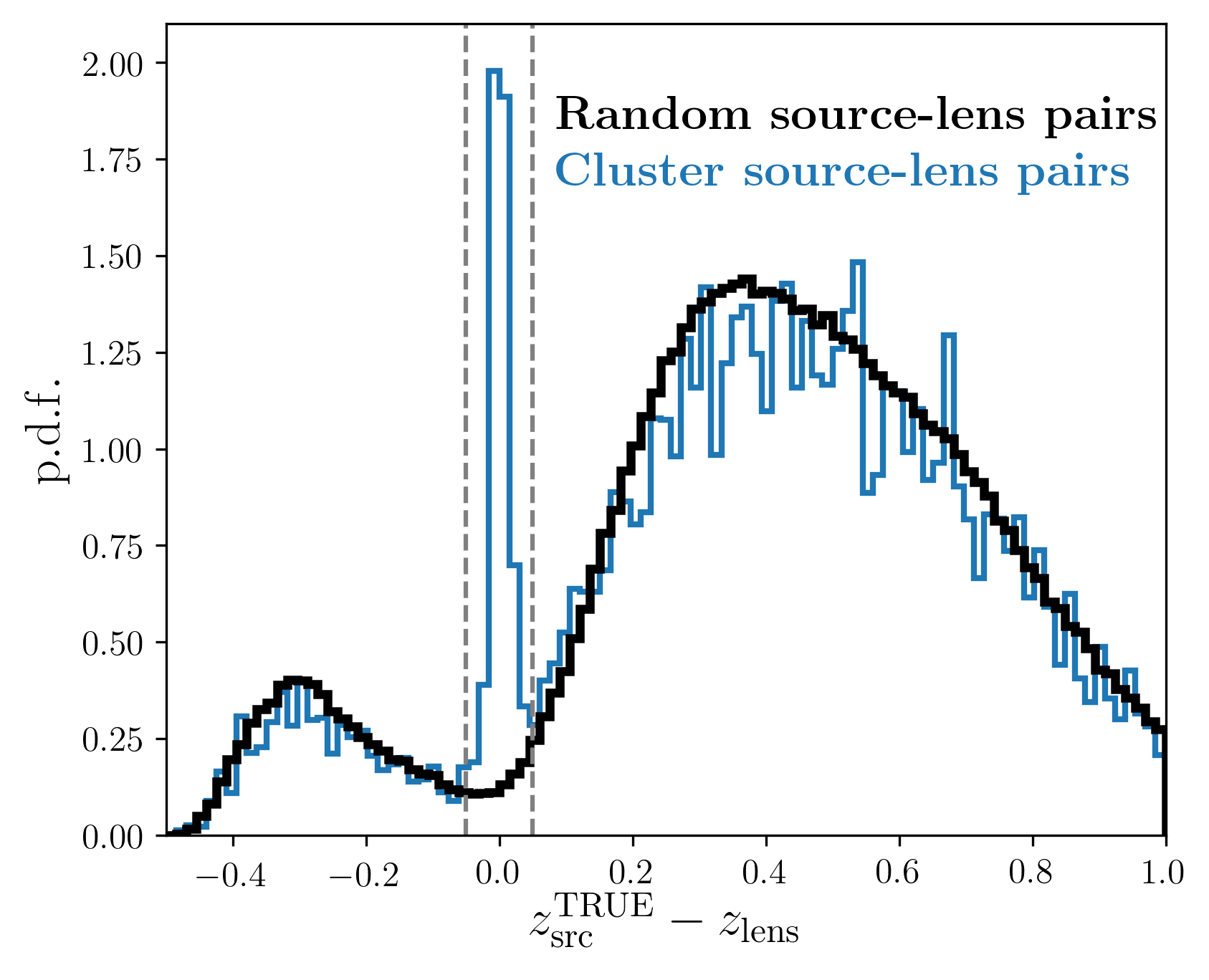}
	\caption{
    Schematic for estimating the \emph{true} cluster member contamination fraction in the \emph{Buzzard} mock observations. The figure shows histograms of the weighted, true redshift separation of source lens-pairs at different radial distances from galaxy clusters with $z\in[0.35; \, 0.5)$ and $\lambda\in[30;\,45)$. 
\emph{Blue:} source-lens pairs at low radial scales around clusters ($R<0.78\, Mpc$).     
\emph{Black:} source--lens pairs around redMaPPer selected random points in the same radial range. 
\emph{Gray dashed:} $\Delta z=\pm0.05$ vicinity of the cluster redshift.
    \label{fig:true_schematic}
    }
\end{figure}


We calculate the true contamination as the \emph{excess} rate for galaxies to be located within the immediate $\mathcal{S}_j \equiv [z_j -\Delta z;\; z_j+\Delta z]$ vicinity of the clusters, defined via
\begin{equation}
	\label{eq:true_boost}
	f_{\rm cl}^\mathrm{true}\equiv \frac{\sum\limits_j^{N_{c}} \sum\limits_{z_{s,i} \in \mathcal{S}_j} \omega_{i,j}  \Sigma'^{-1}_{\rm crit;i,j}}{\sum\limits_{j}^{N_{c}} \sum\limits_{i}^{N_s} \omega_{i,j}  \Sigma'^{-1}_{\rm crit;i,j}} 
- \frac{\sum\limits_l^{N_{r}} \sum\limits_{z_{s,i} \in \mathcal{S}_l} \omega_{i,j}  \Sigma'^{-1}_{\rm crit;i,l}}{\sum\limits_{l}^{N_{r}} \sum\limits_{i}^{N_s} \omega_{i,j}  \Sigma'^{-1}_{\rm crit;i,l}}\,,
\end{equation}
where $N_c$ refers to the number of clusters, $N_r$ to the number of random points, and $N_s$ to the number of source galaxies, while $\omega_{i,j}$ is the lensing weight associated with the source-lens pairs defined in \autoref{eq:updated_weights}. $\mathcal{S}_j$ and $\mathcal{S}_l$ refer to the immediate true redshift vicinities of clusters and random points respectively.
That is, $f_{\rm cl}^{\mathrm{true}}$ is the probability of finding a galaxy within the redshift range $\mathcal{S}$ around the clusters, minus the same probability for random lines-of-sight, where the second term we obtain by saving source-lens pairs around \redmapper\ random points.  This is equivalent to a cylindrical selection of contaminating galaxies, which is motivated by the fact that the contamination originates not only from physically bound galaxies, but also from galaxies in the extended correlated structures.

\autoref{fig:true_schematic} illustrates the above approach. It is clear that a large fraction of source-lens pairs near cluster centers (\emph{blue}) actually lie at the cluster redshift. Comparing this with the distribution of galaxies around random points (\emph{black}), the contamination rate is taken as the excess area under the curve within the $\pm\Delta z $ (\emph{dashed}) region. Based on \autoref{fig:true_schematic}, we adopt $\Delta z = 0.05$ as our fiducial redshift width for the purposes of computing the true contamination rate.
The resulting $f_\mathrm{cl}^\mathrm{true}$ profiles are shown on \autoref{fig:boost_profiles_buzzard} as the blue shaded regions, where the $1\sigma$ uncertainties are estimated from jackknife resampling using the same approach as in  \autoref{sec:buzzard_pz}.

\begin{figure}
\includegraphics[width=\linewidth]{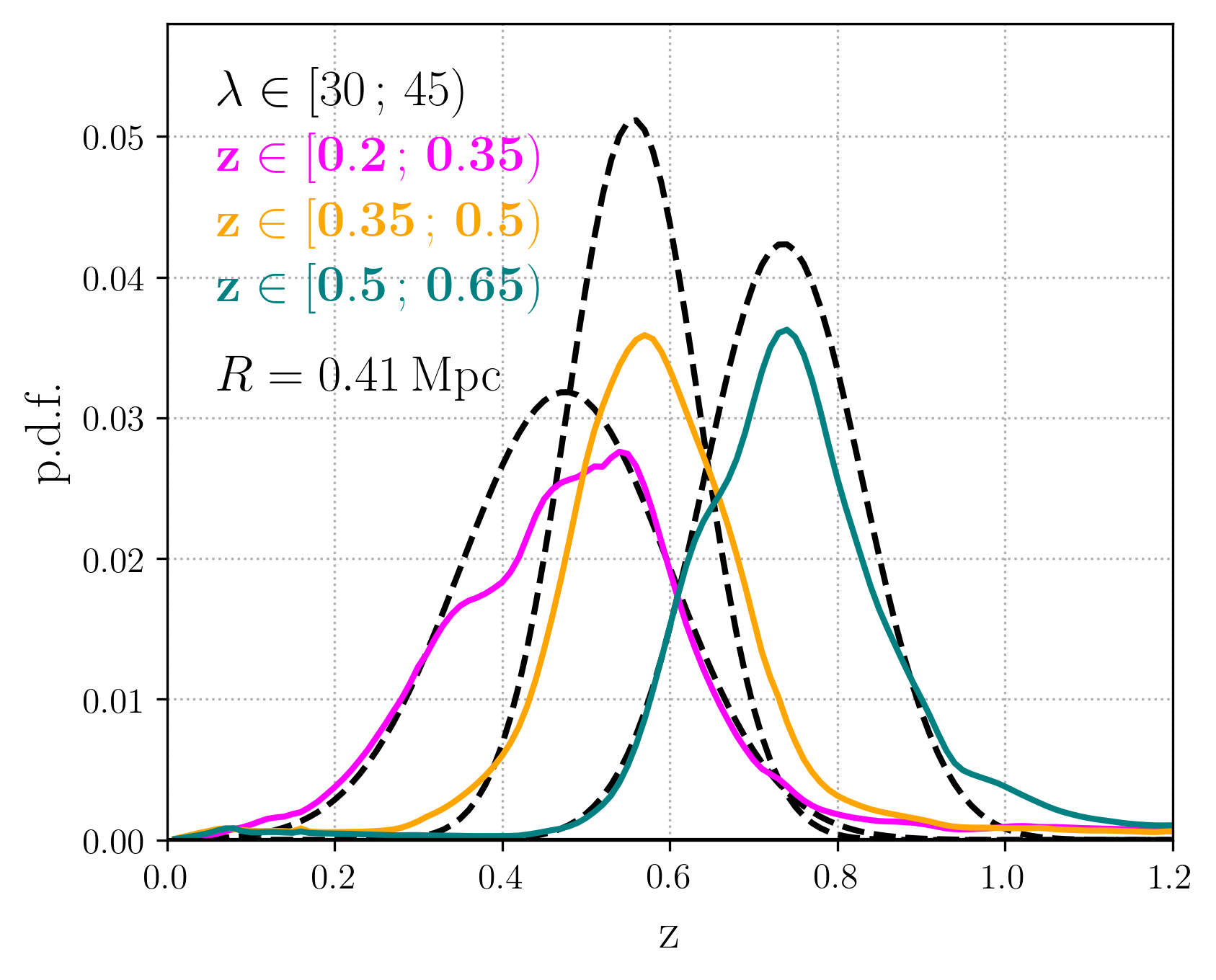}
	\caption{
    Comparison of the Gaussian $P(z)$ component model with the actual $P(z)$-s of likely cluster galaxies in the \emph{Buzzard} mock observations. 
\emph{Colored curves:} $P(z)$ of source galaxies within $\pm\Delta z^\mathrm{TRUE}$ of the clusters, in different cluster redshift bins, but at the same richness and radial bin.
\emph{Dashed curves:} Best fit curves of the Gaussian cluster component model $P_\mathrm{memb}(z)$. Note that the shown P(z)-s are normalized for $z\in[0;\, 3.5]$.
    \label{fig:pz_gauss}
    }
\end{figure}



\subsubsection{Validity of Gaussian cluster model}
\label{sec:member_pz}

With the formalism introduced in \autoref{sec:buzzard_true}, we can compute the average, weighted, photometric redshift $P(z)$ for likely cluster galaxies which are located within the $\pm\Delta z^\mathrm{TRUE}$ vicinities of clusters. As shown on \autoref{fig:pz_gauss} these $P(z)$-s have a prominent peak located slightly above the cluster redshift range. Cluster galaxy $P(z)$-s also possess a long tail extending up to high redshifts. This is an intrinsic feature of photometric redshift estimation, as some cluster galaxies have spectral types that exhibit these types of degeneracies.

\autoref{fig:pz_gauss} also shows the best fit Gaussian cluster component models $P_\mathrm{memb}(z)$. These are obtained from the decomposition method in \autoref{sec:buzzard_pz} and are not informed of the true cluster member $P(z)$-s. Due to the chosen analytic form of $P_\mathrm{memb}(z)$, the long high redshift tail of the actual cluster galaxy $P(z)$-s can not be recovered, which results in the apparent offset of the (normalized) p.d.f-s on \autoref{fig:pz_gauss}. For the shown samples the best fit Gaussian contains within $1\sigma$  57\%, 47\% and 56\% of probability of the actual $P(z)$ of cluster members. Nevertheless, the Gaussians recover the approximate position and width of the peaks, and possesses fewer degrees of freedom than alternative high-skewness models.

\subsubsection{Discussion of simulation benchmarks}
\label{sec:buzzard_summary}


\autoref{fig:boost_profiles_buzzard} compares the boost factor profiles estimated from $P(z)$ decomposition from simulations as described in \autoref{sec:buzzard_pz}, with the actual cluster member contamination rate calculated in \autoref{sec:buzzard_true}.  We find that our estimated boost factors are in excellent agreement with the true member contamination rates in the \emph{Buzzard} mock simulations. The uncertainties shown in \autoref{fig:boost_profiles_buzzard} are estimated via Jackknife resampling, and do not incorporate \emph{systematic} uncertainties. Hence we estimate this systematic uncertainty by requiring consistency between the true and estimated $f_\mathrm{cl}$ profiles across all parameter bins with $N_\mathrm{clust}>50$. Via this approach we find a global relative systematic uncertainty of $\delta_{sys} < 1$ percent across different richness -- redshift selections, where the total covariance is given by $\mathsf{C}_{i,j} = \mathsf{C}^{JK}_{i,j} +  \mathbf{\delta}_{i,j} \cdot f_{\mathrm{cl};\, i}^2  \cdot \delta_{sys}^2$

We note that the simulated galaxy catalogs include the effects of magnification with the typical angular resolution of 0.6 arcminutes, corresponding to approximately $0.15$, $0.2$ and $0.24$ $\mathrm{Mpc}$ in the different redshift bins. While this low resolution allows for only weak constraints, the good agreement between the estimated and true $f_\mathrm{cl}$ profiles indicate that magnification does not play a significant role in the resolved radial ranges.

The purpose of this simulation benchmark is to test how well the $P(z)$ decomposition predictions match the contamination within the simulation, not to extrapolate for the real DES data.  Thus we do not require full realism from the simulated environment. 
Nevertheless, the simulated color distribution of galaxies has been studied by \cite{DeRose2018} in a setup nearly identical to version 1.3  of Buzzard used in the present study. They found the simulated galaxy properties to be broadly consistent with reality except for a slight systematic shift on the color of the blue cloud. We do not anticipate that this manifests in a qualitative difference on the performance of the boost factor estimator compared to real data.
While the abundances, radial profiles, and color properties of cluster galaxies in the simulation may be slightly different from reality, we expect Buzzard to be qualitatively similar to the real DES Y1 data. Hence we take the excellent performance of the $P(z)$ decomposition in this setting as a strong motivation for its applicability for real observations.

\subsection{Analysis on DES Y1 Data}
\label{sec:desy1_tests}

\begin{figure*}
	\includegraphics[width=\linewidth]{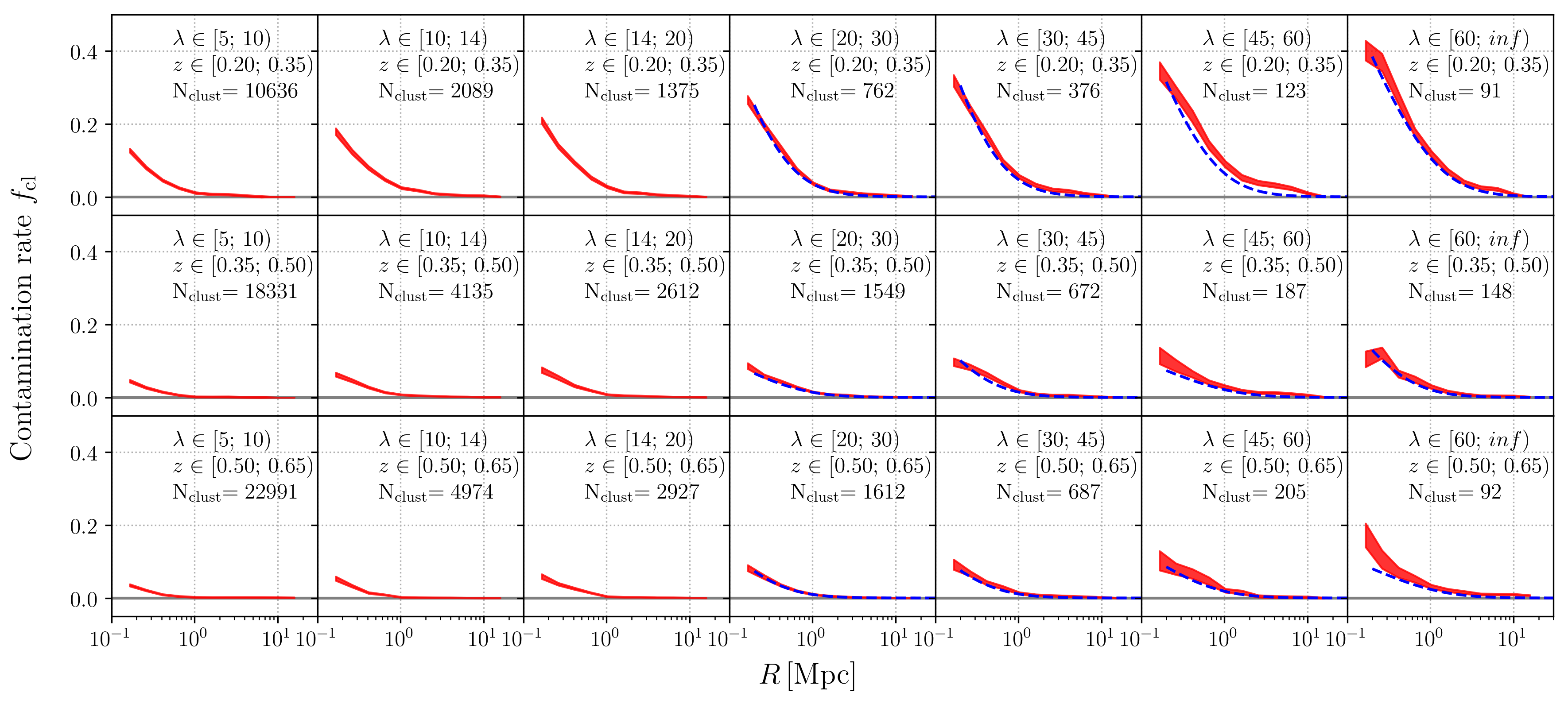}
	\protect{\caption{
    Cluster member contamination measured in the DES Y1 data for the various bins of clusters in richness $\lambda$ and redshift $z$ defined in \autoref{sec:desy1_estimator}. \emph{Red:} contamination profiles calculated via  $P(z)$ decomposition using \bpz\ redshift estimates. \emph{Blue:} Boost factor model from \autoref{sec:boost_model}. The curves correspond to the best fit parameters found by \citet{rmy1} from a likelihood optimization performed jointly with the cluster mass reconstruction.
    \label{fig:desy1_profiles}}
        }
\end{figure*}

In this section we apply the $P(z)$ decomposition method to DES Y1 data, following the exact measurement setup presented in \cite{rmy1}. The structure of this section is the following: in \autoref{sec:desy1data} we present the relevant parts of the DES Y1 dataset relating to the galaxy cluster catalog and the weak lensing source galaxies, in \autoref{sec:desy1_estimator} we derive the form of the necessary boost factor correction, while in \autoref{sec:iterative_test} present a simple test on the robustness of our contamination model, and finally in \autoref{sec:corr_comparison} we compare with the alternative method of correlation based boost factor estimate.

\subsubsection{The DES Y1 dataset}
\label{sec:desy1data}

The DES Y1 observations cover approximately 1800 deg$^2$ of the southern sky in \emph{g,r,i,z} bands. These observations are processed via a variety of photometric data reduction steps into the Y1 GOLD catalog \citep{Y1gold} which is the main science quality catalog of DES. Using the fiducial multi-epoch, multi-object fitting algorithm (MOF) DES finds the 10$\sigma$ limiting magnitudes of this dataset for $2"$ apertures to be $g\approx23.7$,  $r\approx23.5$, $i\approx22.9$ and $z\approx22.2$. Based on these observations \cite{rmy1} defined a locally volume limited catalog of galaxy clusters identified via the redMaPPer algorithm. In the Y1 footprint the average MOF limiting magnitude is deep enough to detect a $0.2\;L^*$  galaxy up to $z\approx0.7$, thereby setting the maximum depth of the volume limited cluster sample.

Approximately 1500 deg$^2$ of this catalog is further processed by the \metacal\ algorithm \citep{HuffMETA,SheldonMETA} to define a source galaxy sample \citep{Y1shape}. This source galaxy catalog consists of an ellipticity estimate $e_i$ for each galaxy, along with ancillary quantities used to perform the bias calibration via the \emph{response} $\mathsf{R} = \mathsf{R}_\gamma  + \mathsf{R}_\mathrm{sel}$ of the ellipticity estimates to shear and the source galaxy selection function respectively.

Photometric redshift $P(z)$-s are calculated via the \bpz\ template based algorithm \citep{Y1pz}. Two separate redshift estimates are derived: one based on the MOF-based galaxy colors listed in the GOLD catalog, and a second based on the photometric model obtained from \metacal. 
This second \metacal\ based redshift estimate is required to properly account for the selection response correction, however is found to have greater scatter compared to to the MOF based redshift estimates. For this reason \cite{rmy1} opted to use the \metacal\ estimates only in selecting and weighting source-lens pairs. \cite{Y1pz} found these redshift estimates to be mildly biased in the mean redshift. Since the $P(z)$ decomposition method is only sensitive to the relative shape of the $P(z)$-s, we do not expect the impact of this bias to be significant.

 
\subsubsection{Contamination estimator for DES Y1-like data}
\label{sec:desy1_estimator}
\label{app:boostpzweights}


In this section we derive the required boost factor correction for the $\Delta\Sigma$ estimator employed for the DES Y1 analysis. For this, \autoref{eq:ds_buzzard} and \autoref{eq:updated_weights} are replaced with:
%
\begin{equation}
\widetilde{\Delta\Sigma}=\frac{\sum \hat{\omega}_{i,j} e_{\rm T;i}}{\sum \hat{\omega}_{i,j} \mathsf{R}^{\rm T}_{\gamma;\,i,j} \Sigma'^{-1}_{\rm crit; \,i,j}} \; .
\end{equation}
where
\begin{equation}
	\label{eq:y1_weights}
	\hat{\omega}_{i,j} \equiv \Sigma_{{\rm crit}}^{-1}\left(z_{{\rm l}_j}, \langle z^\mathrm{MCAL}_{{\rm s}_i}\rangle\right)\;
    \mathrm{if}\; \langle z^\mathrm{MCAL}_{{\rm s}_i}\rangle > z_{{\rm l}_j} + 0.1\,,
\end{equation}
which is the general form of the estimator.
Here we neglected the selection response term, which \cite{rmy1} found to be subdominant compared to the shear response $\mathsf{R}^T_\gamma$, where the superscript refers to the response matrix rotated into the tangential frame.
In the above estimator the weighting and selection is performed based on the mean \metacal\ based redshift estimates $\langle z^\mathrm{MCAL}_{\mathrm{s}_i} \rangle$, while $\Sigma'^{-1}_{\rm crit; \,i,j}$ is calculated using a random draw from the MOF-based redshift $P(z)$. 

Following \autoref{eq:split} and \autoref{eq:f_clust} we find the contamination rate to be:
\begin{equation}
\label{eq:desy1_fcl}
f_\mathrm{cl} = \frac{\sum_{cl} \hat{\omega}_{i,j} \mathsf{R}^{\rm T}_{\rm \gamma;\,i,j} \Sigma'^{-1}_{\rm crit; \,i,j}}{\sum \hat{\omega}_{i,j} \mathsf{R}^{\rm T}_{\rm\gamma;\,i,j} \Sigma'^{-1}_{\rm crit; \,i,j}}\,.
\end{equation}
We perform the $P(z)$ decomposition in a setup identical to \autoref{sec:buzzard_pz}, but using weights according to \autoref{eq:desy1_fcl}, and make use of a randomly selected, representative subsample of the source-lens pairs from \cite{rmy1}. 
The detailed description of our results is presented in \autoref{sec:desy1res}, while the boost profiles themselves are shown on \autoref{fig:desy1_profiles}.



\subsubsection{Sensitivity to background component choice}
\label{sec:iterative_test}

The performance of the $P(z)$  decomposition method is dependent on how well the ansatz for the background component resemble the p.d.f. of actual background galaxies. Furthermore, the average $P(z)$ estimated for a galaxy sample may contain minor features (e.g. wiggles and peaks) which depend on the internal setup of the photometric redshift algorithm (e.g. distribution of templates within the \bpz\ algorithm), and do not themselves relate to the physical distribution of galaxies \citep{BonnettPhotoz2015, rmsva}. Differences originating from these non-physical reasons may also impact the robustness of the contamination estimates.

We test the self-consistency of the decomposition method and its sensitivity to the minor features in the estimated $P(z)$-s by extending the fiducial $P(z)$ decomposition analysis with a second step. In this second step the reference $P(z)$ model is updated from the ``field'' $P(z)$ 
to the observed $P(z)$ at $R\approx 1$ Mpc minus the Gaussian cluster model found in the previous step. The $f_\mathrm{cl}$ fit is then repeated with this new reference $P(z)$ component, while keeping the position and width of the Gaussian cluster model component unchanged. 
The resulting boost factor profiles are shown on \autoref{fig:destests}, overlaid with the fiducial boost factor profiles. The two iterations agree very well, and following the approach used in \autoref{sec:buzzard_summary} we  estimate a relative systematic uncertainty  of $<1$ percent, 
motivating that the  choice for the background component propagates to only a negligible difference in the final contamination profiles.

\subsubsection{Comparison with correlation based boost factors}
\label{sec:corr_comparison}

\begin{figure*}
\includegraphics[width=\linewidth]{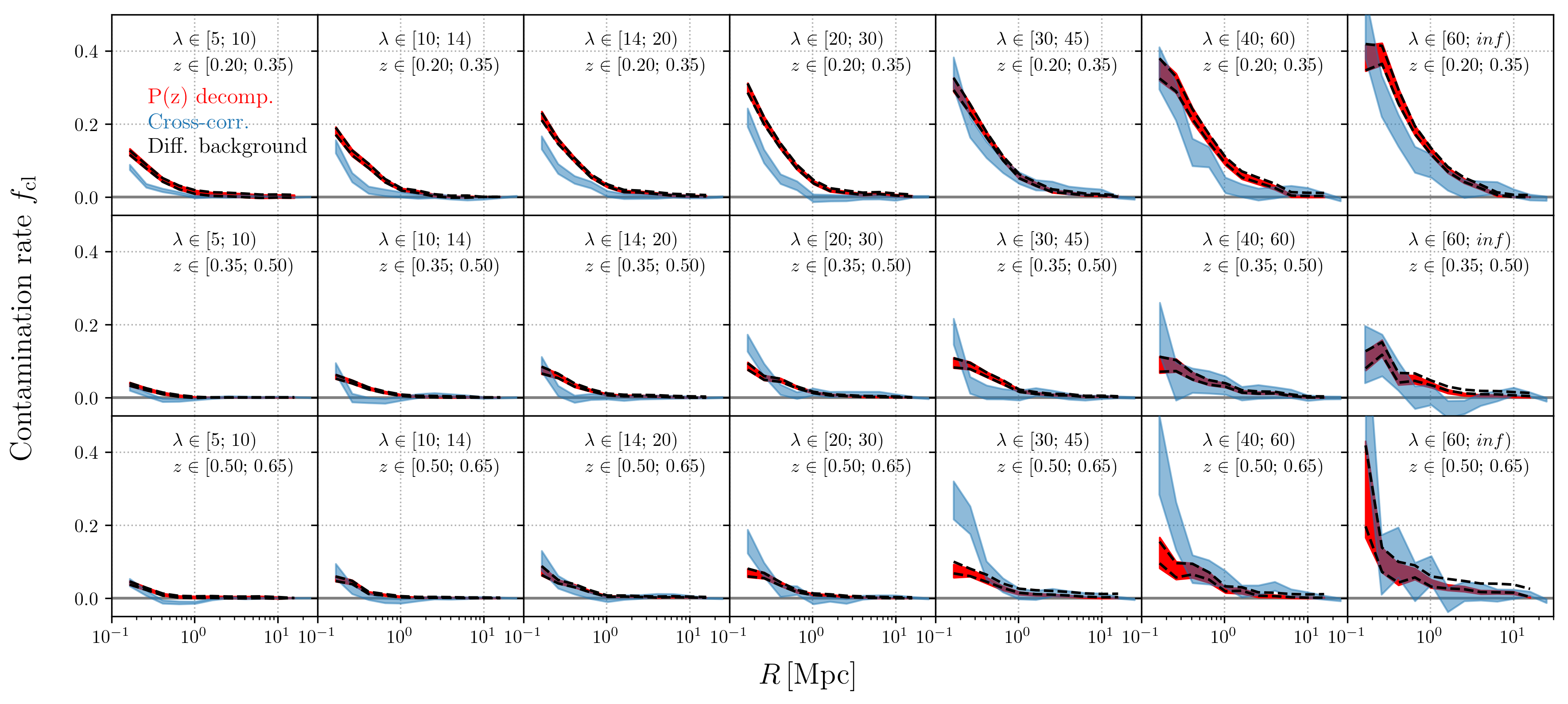}
	\caption{
    Comparison of the $P(z)$ decomposition based contamination estimates \emph{(red)} with the cross-correlation based estimates \emph{(blue)} across the different richness and redshift bins. \emph{Black:} results from the background sensitivity test. The \emph{black} background sensitivity results are found to agree very well with the fiducial red curves, while the \emph{blue} correlation based estimates appear to be globally biased low.
    \label{fig:destests}
    }
\end{figure*}

An alternative way for estimating boost factors is via the angular clustering of source galaxies around clusters, as only the contaminating galaxies are correlated with the cluster \citep{Sheldon04.1, Applegate2014, Hoekstra2015, Simet2017,Leauthaud17}. We calculate this correlation function via the estimator:
\begin{equation}
Corr = \frac{N_{R}}{N_{D}} \cdot \frac{DD}{RR} - 1\,,
\end{equation}
where $DD$ and $RR$ are defined as $\sum \omega \mathsf{R}^{\rm T}_{\gamma} \Sigma'^{-1}_{\rm crit}$ around redMaPPer clusters and  random points respectively, while $N_D$ refers to the number of clusters, and $N_R$ to the number of random points \citep{Landy93}. The results of this measurement are shown on \autoref{fig:destests}. 
The correlation function estimates are, for many cluster samples, preferentially lower than the $P(z)$ decomposition estimates, especially at the two lower redshift selections. 

This can be understood as clusters impacting the spatial distribution of source galaxies in ways other than contamination by cluster galaxies: e.g. the density and blending of cluster members may lead to a bias against selecting sources near clusters \citep[][]{Simet2015, Leauthaud17, Y1shape}, which can explain the preferential lower  estimates. Such effects cannot be captured by random points, as they relate to the presence of the cluster in the line-of-sight, and are not well characterized for the DES Y1 \metacal\ shear catalogs. By contrast, the $P(z)$ decomposition method is insensitive to color-agnostic fluctuations in the source selection, and to the number density profile of source galaxies.

\section{Results for DES Y1 data}
\label{sec:desy1res}

\subsection{Boost factor estimates}

We present the contamination rate estimates from applying our method to the DES Y1 data in \autoref{fig:desy1_profiles}. In the present calculation we consider all cluster selections, but note that in \cite{rmy1} only the $\lambda > 20$ clusters enter the determination of the mass--observable relation.  The qualitative behavior of the contamination rate profiles agrees well with the theoretical expectation of decreasing contamination with increasing radius. As expected, the amplitude of the contamination increases with cluster richness. Furthermore, the contamination rates are higher for low-redshift clusters, as for those fainter cluster member galaxies can also be detected, whose photometric redshifts are less accurate.

\begin{figure}
\includegraphics[width=\linewidth]{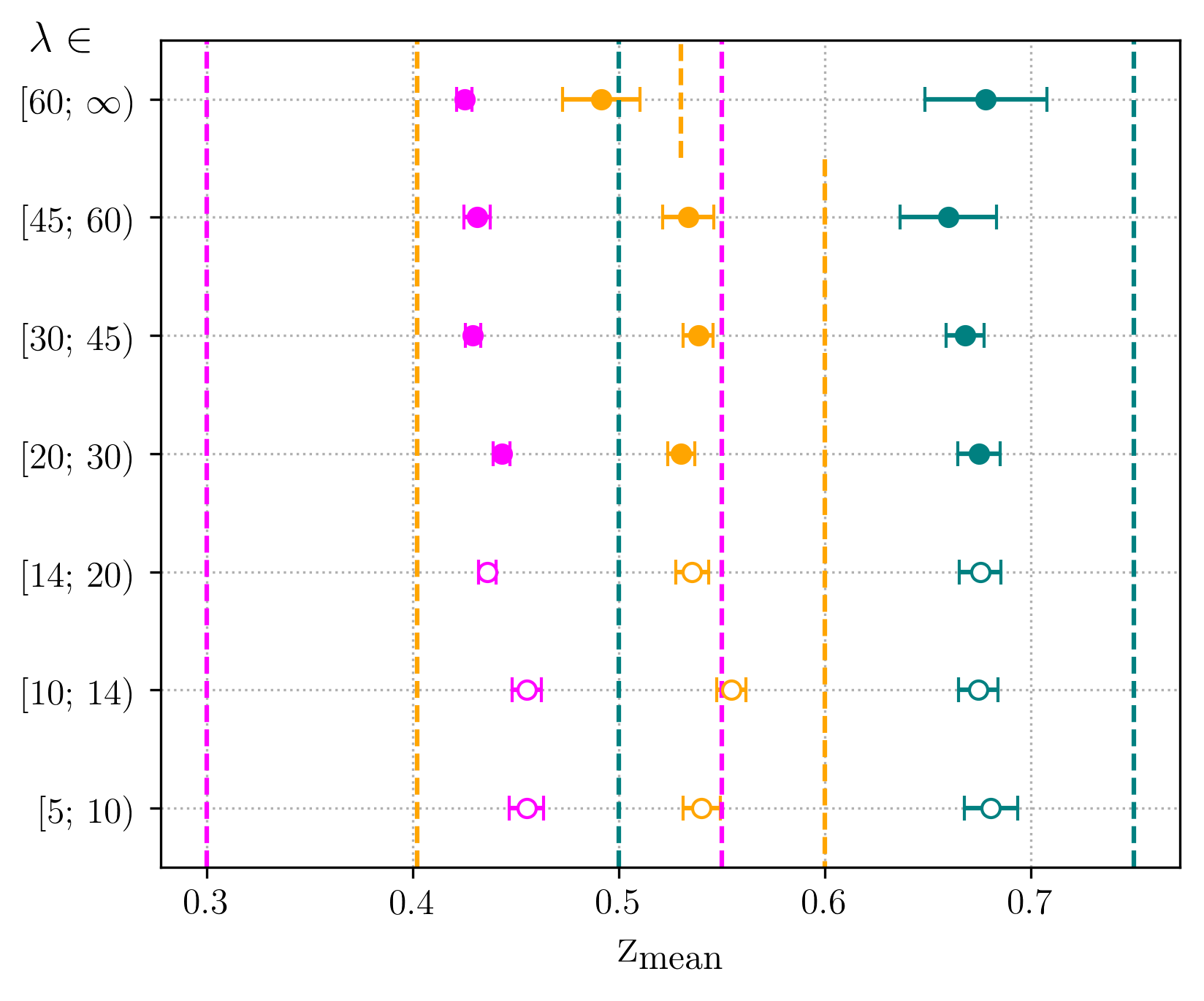}
\includegraphics[width=\linewidth]{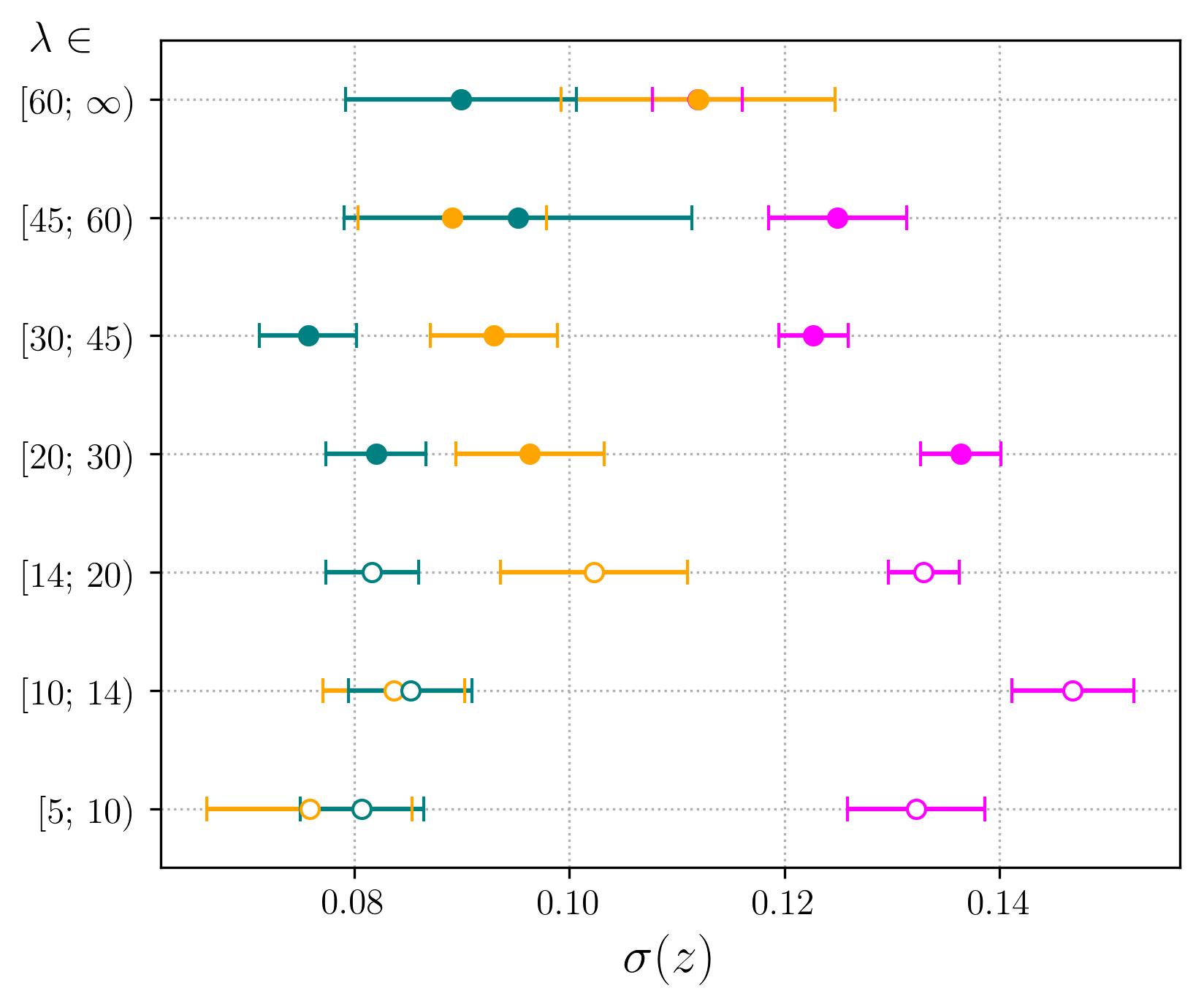}
	\caption{
    Best fit parameters for the Gaussian $P_\mathrm{memb}$ component found for the $P(z)$ decomposition in the DES Y1 data. The colors indicate the various cluster redshift bins: $z\in[0.2, 0.35)$, $z\in[0.35, 0.5)$ and $z\in[0.5, 0.65)$ is denoted by \emph{magenta}, \emph{orange} and \emph{green} respectively.
    \emph{(Top panel:)} mean redshift of the cluster member components. The dashed colored lines indicate the prior range for the mean of the cluster component.
    \emph{(Bottom panel:)} standard deviation of the cluster member components.
    }
    \label{fig:pz_params_desy1}
\end{figure}

We find that the ``peak'' in the $P(z)$ due to contaminating galaxies is very prominent at low radii for all cluster bins, and the presence of this feature is critical for the applicability of the decomposition method. The best fit parameters of the Gaussian $P_\mathrm{memb}(z)$ model are presented in \autoref{fig:pz_params_desy1}, along with the used prior ranges. The means of these best-fit Gaussian $P_\mathrm{memb}(z)$ distributions differ from the redshift ranges of the clusters.  However this is expected from the way source galaxies are selected in the DES analysis: only those cluster member galaxies enter the source selection whose estimated mean redshift scatters towards higher redshifts. 

The contamination rate profiles $f_{cl}(R)$ shown in \autoref{fig:desy1_profiles} can be directly translated into a multiplicative correction factor $\mathcal{B} \equiv (1- f_\mathrm{cl})^{-1}$ necessary for recovering an unbiased estimate on $\Delta\Sigma$  via \autoref{eq:boost_correction}.

\subsection{Analytic boost factor model}
\label{sec:boost_model}

We model the boost factor profile using a Navarro-Frenk-White (NFW) profile \citep{Navarro96.1}:
\begin{equation}
  \label{eq:boost_model}
  \calB_{\rm model}(R) = 1+B_0\frac{1-F(x)}{x^2-1}\,,
\end{equation}
where $x=R/R_s$, and
\begin{align}
  \label{eq:boost_model2}
  F(x) = \left\{
  \begin{array}{lr}
    \frac{\tan^{-1}\sqrt{x^2-1}}{\sqrt{x^2-1}} & : x > 1\\
    1 & : x = 1\\
    \frac{\tanh^{-1}\sqrt{1-x^2}}{\sqrt{1-x^2}} & : x < 1
  \end{array}
  \right.\,.
\end{align}
This model has two free parameters per cluster bin: $B_0$ and $R_s$ which characterize the amplitude and scale radius of the correction profile respectively.

The best fit boost model profiles are overlayed on \autoref{fig:desy1_profiles} to the raw contamination rate estimates. In \citet{rmy1} these fits are performed in a joint likelihood analysis together with the mass profile model and systematic corrections. This way the estimated statistical uncertainty of the boost factors is propagated self-consistently into their final mass constraints. The model parameters are not tied to the mass parameters of clusters to allow for freedom in describing the boost factors. Due to the excellent performance of the $P(z)$ decomposition method in our tests, and as the systematic uncertainty found in a simulated environment was subdominant compared to the Jackknife error estimate, \cite{rmy1} did not assume any additional systematic uncertainty to this source of systematic error. Following the approach used in \autoref{sec:buzzard_summary} we  estimate a relative systematic uncertainty  of $<<1$ percent,

We note that in the Monte Carlo chains run by \cite{rmy1} the $R_s$ and $B_0$ parameters were found to be degenerate, hence the increase in actual contamination does not translate into an obvious increase in $B_0$. However this was found to propagate into only a mild change in the boost factor profile over the studied radial range. The recovered cluster masses were robust against degeneracy in the boost factor model parameters, and are not significantly impacted. Nevertheless, we find that the scale radius of the contamination component is typically at least twice as large as the scale radius of the NFW halo. This is consistent with the expectation that the galaxy distribution of clusters can be described as an NFW distribution with lower (approximately half) concentration than the underlying dark matter halo \citep{Budzynski12}.

\section{Summary and Conclusions}
\label{sec:summary}


In this study we carried out a detailed method validation on the $P(z)$ decomposition cluster member contamination estimation algorithm proposed by \cite{Gruen2014} and \cite{rmsva}. This approach relies on the decomposition of the average redshift $P(z)$-s of source galaxies around galaxy clusters into a cluster member and background component, to obtain an estimate on the relative number of contaminating galaxies which are mistakenly included in the source galaxy catalog.
Since its inception this method has been used by studies ranging from the DES Science Verification cluster mass calibration \citep{rmsva}, to cluster weak lensing studies focusing on the detection of the \emph{splashback}-feature \citep{Chang17}, and to the mass calibration of SPT selected clusters \citep{Stern18}. It also serves as an important constituent of the weak lensing mass calibration on DES Y1 \citep{rmy1}, which will be used in deriving cosmological constraints based on the number counts of optically identified galaxy clusters \citep{y1_cluster_cosmo}.

In order to demonstrate the applicability of  $P(z)$ decomposition based boost factors we performed a series of tests benchmarking various aspects of the approach. We find the following:
\begin{itemize}
\item The method performed  well in a mock survey simulation (\autoref{sec:buzzard_tests}), yielding excellent agreement between the estimated contamination rates and the actual true number of contaminants extracted from the truth catalogs of the simulation.

\item Within the mock analysis we investigated the validity of the Gaussian ansatz for the cluster $P(z)$ component (\autoref{sec:buzzard_true}). We found that it recovers the approximate redshift and width of the peak within the $P(z)$ of the contaminating galaxies.
Furthermore the Gaussian ansatz  did not appreciably bias the estimated contamination.

\item We tested the sensitivity of the contamination estimates to the choice of the background $P(z)$ component on DES Y1 data (\autoref{sec:iterative_test}), and found an excellent agreement between the boost factors derived via the fiducial and alternative background components. 

\item We compared the method with an alternative, transverse correlation based contamination estimate in \autoref{sec:corr_comparison}. We found indications that the alternative method is preferentially underestimating the contaminations, which is likely an imprint of the radial source galaxy selection function. This is expected to impact the decomposition predictions to a lesser extent as it does not make use of the number profile of sources.

\end{itemize}

Excluding galaxies from the source catalog can also reduce the cluster member contamination, however it may also reduce the statistical power of the measurement if the exclusion criteria are too broad. Hence it presents a trade-off in the total error budget between the systematic uncertainty originating from boost factors and statistical uncertainty such as shape noise. However based on the consistency tests presented in this paper, and on the fact that boost factors played a strongly subdominant role in the total error budget of \cite{rmy1} it appears that the $P(z)$ decomposition method is sufficient to provide boost factor estimates for current cluster weak lensing analyses. 
We note that this determination is dependent on the characteristics of the sky survey e.g. depth, area, number of filters. Notably cluster weak lensing studies such as \cite{Medezinski17, HSC18,Miyatake2018} in the ongoing  Hyper Suprime-Cam Survey \citep{HSC_Survey}  favored the approach of trying to excluding cluster member galaxies via color - color or $P(z)$ cuts.

During the DES Y1 analysis we propagated uncertainties by making use of a simple analytic model -- an NFW profile -- to describe the boost factor correction. Previously \cite{rmsva} also used an analytic model, while others  such as \cite{Chang17} and \cite{Stern18} chose to directly use the recovered boost factor profiles in correcting their $\Delta\Sigma$ measurements. While the NFW model was found to be a sufficient description for the analysis of \cite{rmy1}, it is likely that with the increasing precision of future studies more complex boost factor models might become necessary.

We quantified several possible sources of systematic uncertainty impacting the $P(z)$ decomposition method, finding $<1$ percent relative systematic uncertainty based on benchmarks on mock observations, $<1$  percent relative systematic uncertainty originating from the choice of the background $P(z)$ component, and $<1$ percent relative systematic uncertainty from requiring good global agreement between the numerical boost factor estimates and the analytic model. From these contributions we estimate that the decomposition method under optimal circumstances can provide boost factor estimates with approximately $2$ percent relative global systematic uncertainty. However we note that specific circumstances such as the performance of the photometric redshift algorithm, or the source galaxy selection function will impact the accuracy and precision of the $P(z)$ decomposition method.



\section*{Acknowledgments}
This paper has gone through internal review by the DES collaboration. TNV and SS are supported by the SFB-Transregio 33 `The Dark Universe' by the Deutsche Forschungs\-gemeinschaft (DFG) and the DFG cluster of excellence `Origin and Structure of the Universe'. 
JD, DG and RHW received support from the U.S. Department of Energy under
contract number DE-AC02-76SF00515. DG was also supported by Chandra Award
Number GO8-19101A, issued by the Chandra X-ray Observatory Center. 
ER was supported by the DOE grant DE-SC0015975, by the Sloan Foundation, grant FG-2016-6443, and the Cottrell Scholar program of the Research Corporation for Science Advancement. AvdL is supported by the U.S. Department of Energy under Award Number DE-SC0018053.

Part of our computations have been carried out on the computing facilities of the Computational Center for Particle and Astrophysics (C2PAP). A portion of this research was carried out at the Jet Propulsion Laboratory, California Institute of Technology, under a contract with the National Aeronautics and Space Administration. This
research made use of computational resources at SLAC National Accelerator
Laboratory, and the authors thank the SLAC computational team for support.
Some of the computing for this project was performed on the Sherlock
cluster. We would like to thank Stanford University and the Stanford
Research Computing Center for providing computational resources and support
that contributed to these research results. This research also used
resources of the National Energy Research Scientific Computing Center, a
DOE Office of Science User Facility supported by the Office of Science of
the U.S. Department of Energy under Contract No. DE-AC02-05CH11231.

Funding for the DES Projects has been provided by the U.S. Department of Energy, the U.S. National Science Foundation, the Ministry of Science and Education of Spain, 
the Science and Technology Facilities Council of the United Kingdom, the Higher Education Funding Council for England, the National Center for Supercomputing 
Applications at the University of Illinois at Urbana-Champaign, the Kavli Institute of Cosmological Physics at the University of Chicago, 
the Center for Cosmology and Astro-Particle Physics at the Ohio State University,
the Mitchell Institute for Fundamental Physics and Astronomy at Texas A\&M University, Financiadora de Estudos e Projetos, 
Funda{\c c}{\~a}o Carlos Chagas Filho de Amparo {\`a} Pesquisa do Estado do Rio de Janeiro, Conselho Nacional de Desenvolvimento Cient{\'i}fico e Tecnol{\'o}gico and 
the Minist{\'e}rio da Ci{\^e}ncia, Tecnologia e Inova{\c c}{\~a}o, the Deutsche Forschungsgemeinschaft and the Collaborating Institutions in the Dark Energy Survey. 

The Collaborating Institutions are Argonne National Laboratory, the University of California at Santa Cruz, the University of Cambridge, Centro de Investigaciones Energ{\'e}ticas, 
Medioambientales y Tecnol{\'o}gicas-Madrid, the University of Chicago, University College London, the DES-Brazil Consortium, the University of Edinburgh, 
the Eidgen{\"o}ssische Technische Hochschule (ETH) Z{\"u}rich, 
Fermi National Accelerator Laboratory, the University of Illinois at Urbana-Champaign, the Institut de Ci{\`e}ncies de l'Espai (IEEC/CSIC), 
the Institut de F{\'i}sica d'Altes Energies, Lawrence Berkeley National Laboratory, the Ludwig-Maximilians Universit{\"a}t M{\"u}nchen and the associated Excellence Cluster Universe, 
the University of Michigan, the National Optical Astronomy Observatory, the University of Nottingham, The Ohio State University, the University of Pennsylvania, the University of Portsmouth, 
SLAC National Accelerator Laboratory, Stanford University, the University of Sussex, Texas A\&M University, and the OzDES Membership Consortium.

Based in part on observations at Cerro Tololo Inter-American Observatory, National Optical Astronomy Observatory, which is operated by the Association of 
Universities for Research in Astronomy (AURA) under a cooperative agreement with the National Science Foundation.

The DES data management system is supported by the National Science Foundation under Grant Numbers AST-1138766 and AST-1536171.
The DES participants from Spanish institutions are partially supported by MINECO under grants AYA2015-71825, ESP2015-66861, FPA2015-68048, SEV-2016-0588, SEV-2016-0597, and MDM-2015-0509, 
some of which include ERDF funds from the European Union. IFAE is partially funded by the CERCA program of the Generalitat de Catalunya.
Research leading to these results has received funding from the European Research
Council under the European Union's Seventh Framework Program (FP7/2007-2013) including ERC grant agreements 240672, 291329, and 306478.
We  acknowledge support from the Australian Research Council Centre of Excellence for All-sky Astrophysics (CAASTRO), through project number CE110001020, and the Brazilian Instituto Nacional de Ci\^encia
e Tecnologia (INCT) e-Universe (CNPq grant 465376/2014-2).

This manuscript has been authored by Fermi Research Alliance, LLC under Contract No. DE-AC02-07CH11359 with the U.S. Department of Energy, Office of Science, Office of High Energy Physics. The United States Government retains and the publisher, by accepting the article for publication, acknowledges that the United States Government retains a non-exclusive, paid-up, irrevocable, world-wide license to publish or reproduce the published form of this manuscript, or allow others to do so, for United States Government purposes.

\bibliographystyle{mn2e_adsurl}
\bibliography{apj-jour,astroref}

\section*{Affiliations}

$^{1}$ Max Planck Institute for Extraterrestrial Physics, Giessenbachstrasse, 85748 Garching, Germany\\
$^{2}$ Universit\"ats-Sternwarte, Fakult\"at f\"ur Physik, Ludwig-Maximilians Universit\"at M\"unchen, Scheinerstr. 1, 81679 M\"unchen, Germany\\
$^{3}$ Department of Physics, Stanford University, 382 Via Pueblo Mall, Stanford, CA 94305, USA\\
$^{4}$ Kavli Institute for Particle Astrophysics \& Cosmology, P. O. Box 2450, Stanford University, Stanford, CA 94305, USA\\
$^{5}$ SLAC National Accelerator Laboratory, Menlo Park, CA 94025, USA\\
$^{6}$  Brookhaven National Laboratory, Bldg 510, Upton, NY 11973, USA\\
$^{7}$ Department of Physics, University of Arizona, Tucson, AZ 85721, USA\\
$^{8}$ Center for Cosmology and Astro-Particle Physics, The Ohio State University, Columbus, OH 43210, USA\\
$^{9}$ Department of Physics, The Ohio State University, Columbus, OH 43210, USA\\
$^{10}$ Department of Astrophysical Sciences, Princeton University, 4 Ivy Lane, Princeton, NJ 08544\\
$^{11}$ Jet Propulsion Laboratory, California Institute of Technology, 4800 Oak Grove Dr., Pasadena, CA 91109, USA\\
$^{12}$ University of California, Riverside, 900 University Avenue, Riverside, CA 92521, USA\\
$^{13}$ Department of Physics and Astronomy, Stony Brook University, Stony Brook, NY 11794, USA\\
$^{14}$ Fermi National Accelerator Laboratory, P. O. Box 500, Batavia, IL 60510, USA\\
$^{15}$ Institute of Cosmology and Gravitation, University of Portsmouth, Portsmouth, PO1 3FX, UK\\
$^{16}$ CNRS, UMR 7095, Institut d'Astrophysique de Paris, F-75014, Paris, France\\
$^{17}$ Sorbonne Universit\'es, UPMC Univ Paris 06, UMR 7095, Institut d'Astrophysique de Paris, F-75014, Paris, France\\
$^{18}$ Department of Physics \& Astronomy, University College London, Gower Street, London, WC1E 6BT, UK\\
$^{19}$ Centro de Investigaciones Energ\'eticas, Medioambientales y Tecnol\'ogicas (CIEMAT), Madrid, Spain\\
$^{20}$ Laborat\'orio Interinstitucional de e-Astronomia - LIneA, Rua Gal. Jos\'e Cristino 77, Rio de Janeiro, RJ - 20921-400, Brazil\\
$^{21}$ Department of Astronomy, University of Illinois at Urbana-Champaign, 1002 W. Green Street, Urbana, IL 61801, USA\\
$^{22}$ National Center for Supercomputing Applications, 1205 West Clark St., Urbana, IL 61801, USA\\
$^{23}$ Institut de F\'{\i}sica d'Altes Energies (IFAE), The Barcelona Institute of Science and Technology, Campus UAB, 08193 Bellaterra (Barcelona) Spain\\
$^{24}$ Department of Physics and Astronomy, University of Pennsylvania, Philadelphia, PA 19104, USA\\
$^{25}$ Observat\'orio Nacional, Rua Gal. Jos\'e Cristino 77, Rio de Janeiro, RJ - 20921-400, Brazil\\
$^{26}$ Department of Physics, IIT Hyderabad, Kandi, Telangana 502285, India\\
$^{27}$ Excellence Cluster Universe, Boltzmannstr.\ 2, 85748 Garching, Germany\\
$^{28}$ Faculty of Physics, Ludwig-Maximilians-Universit\"at, Scheinerstr. 1, 81679 Munich, Germany\\
$^{29}$ Department of Astronomy, University of Michigan, Ann Arbor, MI 48109, USA\\
$^{30}$ Department of Physics, University of Michigan, Ann Arbor, MI 48109, USA\\
$^{31}$ Institut d'Estudis Espacials de Catalunya (IEEC), 08034 Barcelona, Spain\\
$^{32}$ Institute of Space Sciences (ICE, CSIC),  Campus UAB, Carrer de Can Magrans, s/n,  08193 Barcelona, Spain\\
$^{33}$ Kavli Institute for Cosmological Physics, University of Chicago, Chicago, IL 60637, USA\\
$^{34}$ Instituto de Fisica Teorica UAM/CSIC, Universidad Autonoma de Madrid, 28049 Madrid, Spain\\
$^{35}$ Department of Physics, ETH Zurich, Wolfgang-Pauli-Strasse 16, CH-8093 Zurich, Switzerland\\
$^{36}$ Santa Cruz Institute for Particle Physics, Santa Cruz, CA 95064, USA\\
$^{37}$ Harvard-Smithsonian Center for Astrophysics, Cambridge, MA 02138, USA\\
$^{38}$ Australian Astronomical Optics, Macquarie University, North Ryde, NSW 2113, Australia\\
$^{39}$ Departamento de F\'isica Matem\'atica, Instituto de F\'isica, Universidade de S\~ao Paulo, CP 66318, S\~ao Paulo, SP, 05314-970, Brazil\\
$^{40}$ George P. and Cynthia Woods Mitchell Institute for Fundamental Physics and Astronomy, and Department of Physics and Astronomy, Texas A\&M University, College Station, TX 77843,  USA\\
$^{41}$ Department of Astrophysical Sciences, Princeton University, Peyton Hall, Princeton, NJ 08544, USA\\
$^{42}$ Instituci\'o Catalana de Recerca i Estudis Avan\c{c}ats, E-08010 Barcelona, Spain\\
$^{43}$ Department of Physics and Astronomy, Pevensey Building, University of Sussex, Brighton, BN1 9QH, UK\\
$^{44}$ School of Physics and Astronomy, University of Southampton,  Southampton, SO17 1BJ, UK\\
$^{45}$ Instituto de F\'isica Gleb Wataghin, Universidade Estadual de Campinas, 13083-859, Campinas, SP, Brazil\\
$^{46}$ Computer Science and Mathematics Division, Oak Ridge National Laboratory, Oak Ridge, TN 37831\\




\label{lastpage}
\end{document}